\documentclass[]{aa}
\pdfoutput=1
\usepackage{amsmath}
\usepackage{graphicx}
\usepackage[varg]{txfonts}
\usepackage{natbib}
\bibpunct{(}{)}{;}{a}{}{,}
\usepackage{epstopdf}
\usepackage{array}

\usepackage[usenames,dvipsnames,svgnames,table]{xcolor}
\usepackage[colorlinks,allcolors=MidnightBlue,]{hyperref} 

\begin{document}  
 
   \title{OGLE-2014-SN-131: A long-rising Type Ibn supernova from a massive progenitor} 
   
\titlerunning{OGLE-2014-SN-131: a long-rising SN~Ibn}
\authorrunning{E. Karamehmetoglu et al.}
   \author{E. Karamehmetoglu \inst{1} \and 
          F. Taddia \inst{1} \and 
          J. Sollerman \inst{1} \and 
          {\L}. Wyrzykowski \inst{2} \and 
          S. Schmidl \inst{3} \and 
          M. Fraser  \inst{4} \and 
          C. Fremling \inst{1} \and
          J. Greiner \inst{5,6} \and
          C. Inserra \inst{7} \and
          Z. Kostrzewa-Rutkowska \inst{2,8,9} \and
          K. Maguire \inst{7} \and
          S. Smartt  \inst{7} \and 
          M. Sullivan \inst{10} \and 
          D. R. Young  \inst{7} 
          }

   \institute{Department of Astronomy, The Oskar Klein Centre, Stockholm University, AlbaNova, 10691 Stockholm, Sweden.\\
            \email{emir.k@astro.su.se}
    \and Warsaw University Astronomical Observatory, Al. Ujazdowskie 4, 00-478 Warszawa, Poland.
    \and Th\"uringer Landessternwarte Tautenburg, Sternwarte 5, 07778 Tautenburg, Germany. 
    \and Institute of Astronomy, University of Cambridge, Madingley Road, Cambridge CB3 0HA, UK.
    \and Max-Planck-Institut f\"ur Extraterrestrische Physik, Giessenbachstrasse 1, 85748 Garching, Germany.
    \and Excellence Cluster Universe, Technische Universit\"at M\"unchen, Boltzmannstrasse 2, 85748 Garching, Germany.
    \and Astrophysics Research Centre, School of Mathematics and Physics, Queens University Belfast, Belfast BT7 1NN, UK. 
    \and SRON Netherlands Institute for Space Research, Sorbonnelaan 2, 3584 CA Utrecht, the Netherlands.
    \and Department of Astrophysics/IMAPP, Radboud University Nijmegen, P.O. Box 9010, 6500 GL Nijmegen, the Netherlands.
    \and Department of Physics and Astronomy, University of Southampton, Southampton, SO17 1BJ, UK. 
                      }

   \date{Received date /
   Accepted date } 

  \abstract
   {Type Ibn supernovae (SNe~Ibn) are thought to be the core-collapse explosions of massive stars whose ejecta interact with He-rich circumstellar material (CSM).}
   {We report the discovery of a SN~Ibn, with the longest rise-time ever observed, OGLE-2014-SN-131. We discuss the potential powering mechanisms and the progenitor nature of this peculiar stripped-envelope (SE), circumstellar-interacting SN.}
   {Optical photometry and spectroscopy were obtained with multiple telescopes including VLT, NTT, and GROND. We compare light curves and spectra with those of other known SNe Ibn and Ibc. CSM velocities are derived from the spectral analysis. The SN light curve is modeled under different assumptions about its powering mechanism (${^{56}}$Ni decay, CSM-interaction, magnetar) in order to estimate the SN progenitor parameters.}
   {OGLE-2014-SN-131 spectroscopically resembles SNe~Ibn such as SN~2010al. Its peak luminosity and post-peak colors are also similar to those of other SNe~Ibn. However, it shows an unprecedentedly long rise-time and a much broader light curve compared to other SNe~Ibn. Its bolometric light curve can be reproduced by magnetar and CSM-interaction models, but not by a $^{56}$Ni-decay powering model.}
   {To explain the unusually long rise-time, the broad light curve, the light curve decline, and the spectra characterized by narrow emission lines, we favor a powering mechanism where the SN ejecta are interacting with a dense CSM. The progenitor of OGLE-2014-SN-131 was likely a Wolf-Rayet star with a mass greater than that of a typical SN Ibn progenitor, which expelled the CSM that the SN is interacting with.}

   \maketitle 

\section{Introduction}  
Normal Type Ib/c supernovae (SNe~Ibc) are powered by the decay of $^{56}$Ni and $^{56}$Co synthesized in the explosion and present in the ejecta. This mechanism explains the luminosity and light curve shape of these transients, which are characterized by bell-shaped light curves with relatively short rise times of typically $15-20$ days. Besides radioactive decay, an additional powering source coming from the spin-down of a young, rapidly-spinning, magnetar was invoked for the peculiar Type Ib SN~2005bf, which had a double-peaked light curve \citep[e.g.,][]{2006ApJ...641.1039F}.
The progenitors of these stripped-envelope (SE) core-collapse (CC) SNe can be either single massive stars stripped by metal-driven winds, or less massive stars in binary systems, stripped by their companions \citep{2009ARA&A..47...63S,2015PASA...32...15Y}. The low ejecta masses deduced from their narrow light curves favor the binary progenitor scenario for the majority of SNe~Ibc  \citep[e.g.,][]{2013MNRAS.434.1098C,2015A&A...574A..60T,2016MNRAS.457..328L}. Pre-explosion detection searches have also noted a lack of very massive progenitor stars for SE SNe \citep{2013MNRAS.436..774E}. 

Beginning with the observations of SN~2006jc \citep{2007Natur.447..829P,2007ApJ...657L.105F,2008MNRAS.389..141M} and other similar SE~SNe with narrow He emission lines, interaction of the SN ejecta with the progenitor circumstellar medium (CSM) was proposed as an additional powering source for SE~SNe. These CSM-interacting SE~SNe are now referred to as SNe~Ibn. They show prominent helium lines, similar to Type Ib SNe, although with narrow emission lines, similar to Type IIn SNe \citep[e.g.,][]{2016MNRAS.456..853P}.   

The first reported SN with strong narrow He lines in emission was SN~1999cq \citep{2000AJ....119.2303M}, and the subsequent discovery of SN~2006jc linked these two objects together and established the definition of the class of SNe~Ibn. SN 2006jc also showed a pre-explosion outburst two years before collapse \citep{2007Natur.447..829P,2007ApJ...657L.105F}. A number of SNe~Ibn have since been discovered, and \citet{griffin2016} recently presented an analysis of the full sample of 22 objects from the literature.

SNe belonging to the Type Ibn class have relatively high peak luminosities ($M_R\sim-17 \text{ to} -20~\text{mag}$), are blue at peak, and their spectra display narrow ($\lesssim$ few $\times 10^3\text{ km s}^{-1}$) He emission lines superimposed on a typical SN~Ibc spectrum \citep{2016MNRAS.456..853P}. Other than the above criteria used to classify them as Type Ibn, the SNe of this class display a variety of observational characteristics. They are most often fast evolving with a short rise to peak brightness ($\lesssim 2~\text{weeks}$) and a subsequent fast decline, but a slowly evolving example, with a post-peak plateau phase, has also been found (OGLE-2012-SN-006; \citealp{2015MNRAS.449.1941P}). Their spectra can show some amount of hydrogen in emission \citep[see e.g.,][]{2007ApJ...657L.105F,2007Natur.447..829P} or He features in absorption as broad as a normal Type Ib \citep{2015MNRAS.453.3649P}, which has led to the suggestion that SNe~Ibn form part of a continuum from SNe~Ib to SNe~IIn \citep{2015MNRAS.449.1921P}. By studying the rise-time and peak magnitude of their light curves, \cite{Moriya2016} have argued that Type Ibn and IIn SNe have similar explosion properties and circumstellar density. However, they also note that a SN Ibn generally declines faster than a SN IIn, and suggest that the mass-loss, which created the CSM
has probably occurred over a shorter timescale. The same rapid decline might also indicate a lower explosion energy for a SN Ibn compared to other SE SNe.

Based on the rapid evolution and typical peak luminosities of their light curves, \citet{griffin2016} argued that SNe Ibn have relatively uniform light curves as compared to other CSM interacting supernovae, specifically compared to Type IIn SNe. This apparent uniformity might imply that SNe Ibn come from a similarly uniform set of progenitors. The progenitors of SNe~Ibn have been proposed to be Wolf-Rayet (WR) stars \citep{2007Natur.447..829P}. This has been inferred from the presence of active star-formation in their environments, the large line velocities of the He-rich CSM, and the similarity of the underlying spectrum to a Type Ic spectrum. In a detailed study, \citet{2016MNRAS.456..853P} have shown that a majority of SNe Ibn fit this characterization. However, the discovery of a single SN Ibn located in an elliptical galaxy has also led to the suggestion that SNe Ibn could be thermonuclear explosions arising from evolved progenitors \citep{2013ApJ...769...39S}.

In this paper, we present a SN~Ibn with an unusually broad light curve and the longest rise-time ever observed: OGLE-2014-SN-131. While spectroscopically identical to other SNe~Ibn, the light curve of OGLE-2014-SN-131 is both broader, and at least 30~days longer rising than the average Type Ibn SN with a well constrained rise time \citep{griffin2016}. By comparison to other SE SNe with broad light curves, and to analytical light curve models, we argue that the photometry of OGLE-2014-SN-131 might be explained if this SN had a higher mass progenitor compared to other SNe Ibn. 

This paper is organized as follows: basic SN information is given in Sect.~\ref{sec:disco}; observations and data reductions are presented in Sect.~\ref{sec:observations}; spectroscopic analysis is shown in Sect.~\ref{sec:spec}. OGLE-2014-SN-131's photometry is presented in Sect.~\ref{sec:phot}; we then undertake simple modeling to investigate the powering mechanism for OGLE-2014-SN-131 in Sect.~\ref{sec:model}; a discussion of our results and the main conclusions are given in Sect.~\ref{sec:discussion} and Sect.~\ref{sec:conclusion}, respectively.
  
\begin{figure}
    \centering
    \includegraphics[width=\linewidth]{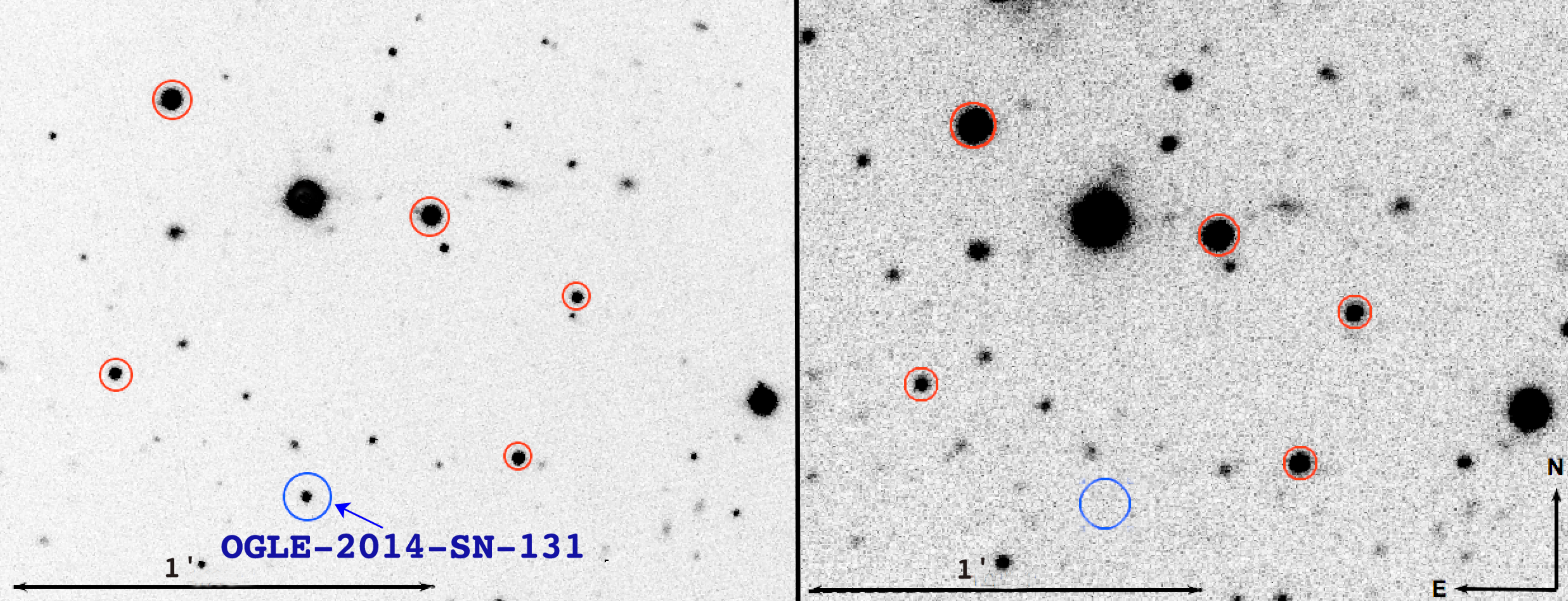}
    \caption{Left: OGLE-2014-SN-131 (blue circle) in a VLT acquisition image taken with an $R-$special filter on November 29.1, 2014 (JD 2456990.08). Right: NTT image from August 1.2, 2016 (JD 2457601.74), showing no visible host at the SN location. Comparison stars in the field from Table \ref{tab:calib} have been marked with red circles in both frames.}
    \label{Fig1}
\end{figure}

\section{OGLE-2014-SN-131 basic information}
\label{sec:disco}
The transient OGLE-2014-SN-131, hereafter OGLE14-131, was discovered in the $I$ band at 19.8~mag on November 11.05 2014 UT by the OGLE-IV Transient Detection System (\citealt{2014ATel.6700....1W}, see Fig.~\ref{Fig1} for a finding chart). Optimized image analysis at the position of the transient was used to obtain $I$-band measurements prior to discovery, with the earliest detection on September 28.2 2014 at 21.3~mag. 
 
OGLE14-131 is located at \(\alpha\) = $01^h14^m00^s.81$ \(\delta\) = $-77\degr06\arcmin16\farcs5$ (J2000.0) with no visible host. The Public ESO Spectroscopic Survey for Transient Objects (PESSTO; \citealp{2015A&A...579A..40S}) obtained a spectrum on November 15 2014, but were unable to classify the SN due to an incorrect initial redshift estimate \citep{2014ATel.6706....1D}. We revise the redshift estimate for OGLE14-131 obtaining z=0.085$\pm$0.002 (see Sect.~\ref{sec:spec}). Based on this redshift, we use a distance modulus of 37.85$\pm$0.05 mag, corresponding to a luminosity distance of 372$\pm$9~Mpc. Here we assumed $H_{0} = 73\,\text{km}\,\text{s}^{-1} \text{Mpc}^{-1}, \Omega_M = 0.27, \Omega_\Lambda = 0.73$ to allow for a direct comparison with other SNe~Ibn whose distance moduli were computed with these cosmological parameters (see table 5 in \citealp{2016MNRAS.456..853P}).

To correct for the Milky Way extinction, we used a color excess of $E(B-V)_{MW}= 0.05$~mag \citep{2011ApJ...737..103S} from NED\footnote{NASA/IPAC Extragalactic Database (NED) is operated by the Jet Propulsion Laboratory, California Institute of Technology, under contract with the National Aeronautics and Space Administration. https://ned.ipac.caltech.edu}, and assumed a standard ($R_V=3.1$) reddening law \citep{1999PASP..111...63F}.

In order to assess the degree of extinction occurring in the host galaxy, we inspected the VLT spectrum (see Sect.~\ref{sec:spec}) taken around maximum light for any signature of narrow \ion{Na}{i}~D in absorption. This feature is traditionally used to gauge host extinction. Since the line was not found, we assume the contribution of the host galaxy to the total reddening to be negligible. 

Additionally, we estimated an upper limit to the host extinction by determining an upper limit to the narrow \ion{Na}{i}~D equivalent width (EW) and applying the empirical relation of \cite{2012MNRAS.426.1465P}. Assuming that the \ion{Na}{i}~D  feature is hidden in the noise of the spectrum at its known wavelength, we assumed the hidden absorption line to be a Gaussian with full width at half maximum (FWHM) equal to the resolution of the spectrum, and the depth equal to the standard deviation of the signal in that region. The EW of this Gaussian corresponds to a host color excess of $E(B-V)_{\rm host}\lesssim0.11$ mag, which is our upper limit. 

Given the first (erroneous) redshift estimate of $z\simeq0.31$ and the broadness of the observed light curve, OGLE14-131 was initially thought to be a super-luminous SN. With our revisions, we classify OGLE14-131 as a SN~Ibn based on strong and narrow \ion{He}{i} emission features dominating over a weak narrow H\(\alpha\) emission line (see Sect.~\ref{sec:spec}). 

To search for a host galaxy, we took $1800$ seconds of $B$-band imaging with the ESO New Technology Telescope (NTT) at La Silla on August 1.2 2016, long after the SN had faded. The field was calibrated using 13 of the in-field comparison stars from Table \ref{tab:grondphot}. Comparison star magnitudes in $g'$ and $r'$ were converted to Vega $B$-band magnitudes using the relation from \citet{2005AJ....130..873J}. We detected no host and estimate a two-sigma upper limit of $24.2$ mag in the $B$ band, which corresponds to an absolute magnitude of $-13.6 \pm 0.05$ using the distance modulus of OGLE14-131. Following the empirical relation in \citet{2004ApJ...613..898T}, the host galaxy of OGLE14-131 must have a metallicity less than $12 + \text{log}(O/H)= 7.78$, which is lower than the average metallicity of the Small Magellanic Cloud \citep{2003A&A...397..487P}. 
 
\section{Observations and data reduction}
\label{sec:observations}
After discovery, OGLE14-131 was photometrically followed by the Optical Gravitational Lensing Experiment (OGLE; \citealp{2014AcA....64..197W}) telescope and later with the Gamma-Ray Burst Optical/Near-Infrared Detector (GROND) \citep{2008PASP..120..405G} on the 2.2m telescope at La Silla (Chile), until its disappearance half a year later. We also obtained spectroscopy with the NTT at La Silla and with the Very Large Telescope (VLT) at Cerro Paranal around maximum light and on the declining part of the light curve.

\subsection{Photometry}
Between September 7 2014 and January 19 2015, the field including OGLE14-131 was observed primarily in the $I$~band with the OGLE telescope \citep{2014AcA....64..197W}. These observations span the pre-discovery, rise, peak, and decline of the SN. The same field was also imaged 16 times in the $V$ band during the decline phase of the light curve. After November 15 2014, we obtained eight epochs of multi-band photometry with GROND \citep{2008PASP..120..405G} in the $g'$, $r'$, $i'$, $z'$, $J$, $H$, $K$ bands, though the SN was never detected in the $H$ and $K$ bands. These observations started shortly after maximum and covered the light curve decline until January 26 2015 (see Tables~\ref{tab:oglephot} and \ref{tab:grondphot}). 

Though there was no visible host in the images, photometry obtained with the OGLE telescope was template subtracted by the standard photometric pipeline of the OGLE IV survey \citep{2014AcA....64..197W}. We did not perform any template subtraction for the GROND photometry. 

GROND data have been reduced in the standard manner using pyraf/IRAF \citep{Tody1993, kkg08, kkg08b}. The optical photometry was measured using point-spread function (PSF) photometry, while the near-infrared (NIR) bands were measured with aperture photometry. The optical photometry was calibrated against comparison stars obtained by observing a Sloan Digital Sky Survery (SDSS) field \citep{2014ApJS..211...17A} under photometric conditions, and calibrated against the primary SDSS standard star network\footnote{http://www.sdss.org}. The NIR data were calibrated against the 2MASS catalog\footnote{This results in typical absolute accuracies  of $\pm$0.03~mag in $g^\prime r^\prime i^\prime z^\prime$  and $\pm$0.05~mag in $JHK_{\rm s}$.}.

The OGLE $I$ and $V$ bands are similar to the standard Johnson-Cousins $I$ and $V$ bands \citep{2015AcA....65....1U}, while the GROND $g'$, $r'$, $z'$ bands are similar to the Sloan filters. The GROND $i'$ band is substantially narrower than the Sloan $i$ band. The magnitudes in the GROND optical bands are provided in the AB system, while the NIR bands are in the Vega system. Details of all GROND filters can be found in \citet{2008PASP..120..405G}.  

The OGLE photometry was calibrated after the image differencing process using the zero-point of the template image, which was obtained from several bright stars in the field. The GROND photometry in the $g'$, $r'$, $i'$, $z'$, and $J$ bands was calibrated using an ensemble of 18 stars in the field . We report the positions and GROND photometry of the calibration stars in Table~\ref{tab:calib}.

\tabcolsep=0.11cm
\begin{table}
\centering
     \scalebox{0.85}{
     \begin{tabular}{lllll}
        \hline
        \noalign{\smallskip}
        Date & Phase & Telescope + Instrument & Range & Resolution \\
        \noalign{\smallskip}
        \hline
        \noalign{\smallskip}
        (y/m/d) & (d) &  & (\AA) & (\AA) \\
        \noalign{\smallskip}
        \hline
        \noalign{\smallskip}
        2014/11/15 & +2.6 &  NTT + EFOSC2 grism 13 & 3661-8815 & 18.2  \\ 
        2014/11/16 & +3.5 &  NTT + EFOSC2 grism 16 & 6006-9786 & 13.4   \\ 
        2014/11/29 & +15.4 & VLT + FORS2 grism 300I & 6150-9261 & 13.0  \\ 
        2014/12/14 & +29.3 & NTT + EFOSC2 grism 11 & 3582-7463 & 13.8 \\ 
        \noalign{\smallskip}
        \hline
     \end{tabular} 
     }
 \caption[]{Log of spectral observations. A slit width of 1\farcs0 was used for every observation. Phases are rest frame days measured from JD of $I-$band maximum, JD 2456973.9. Resolution is the standard value of the given instrumental configuration for a 1\farcs0~slit at the central wavelength. The range reflects the useful range plotted in Fig.~\ref{specseq}.}
 \label{tab:speclog}
\end{table}

\subsection{Spectroscopy}

On three occasions (November 15, 16, and December 14 in 2014), spectra were obtained using the ESO NTT telescope with the EFOSC2 instrument, using grism \#13, \#16, and \#11 respectively as part of a PESSTO monitoring campaign. A slit width of 1\farcs0 was used. We tabulate the instrumental configuration and basic spectral properties in Table.~\ref{tab:speclog}.  

The spectra were reduced with the PESSTO pipeline (Version ntt\_2.6.6) as described in \cite{2015A&A...579A..40S}\footnote{www.pessto.org}. The pipeline provides wavelength calibration using HeAr arc frames, and flux calibration using the spectrum of a spectrophotometric standard star. The spectra are publicly available in Spectroscopic Survey Data Release 3 from the ESO Science Archive Facility\footnote{for access see: http://www.pessto.org/index.py?pageId=58}. 

On November 29 2014, we obtained a spectrum using FORS2 on VLT\footnote{Program ID:294.D-5011(A), PI: \L{}ukasz Wyrzykowski.} equipped with grism 300I and a 1\farcs0 wide slit, (see Table~\ref{tab:speclog}).
The observations were made in the long-slit spectroscopy mode using offsets along the slit, and the data were reduced with the standard recipes for spectroscopic reduction of FORS2 data in ESO's \textit{Reflex} pipeline \citep{2013A&A...559A..96F}. The pipeline allows for sky subtraction, wavelength calibration using a HeAr arc lamp, and flux calibration with a spectrophotometric standard star.

The spectra are available via WISeREP \citep{2012PASP..124..668Y}. 
 
   \begin{figure}
   \centering
   \includegraphics[width=\linewidth]{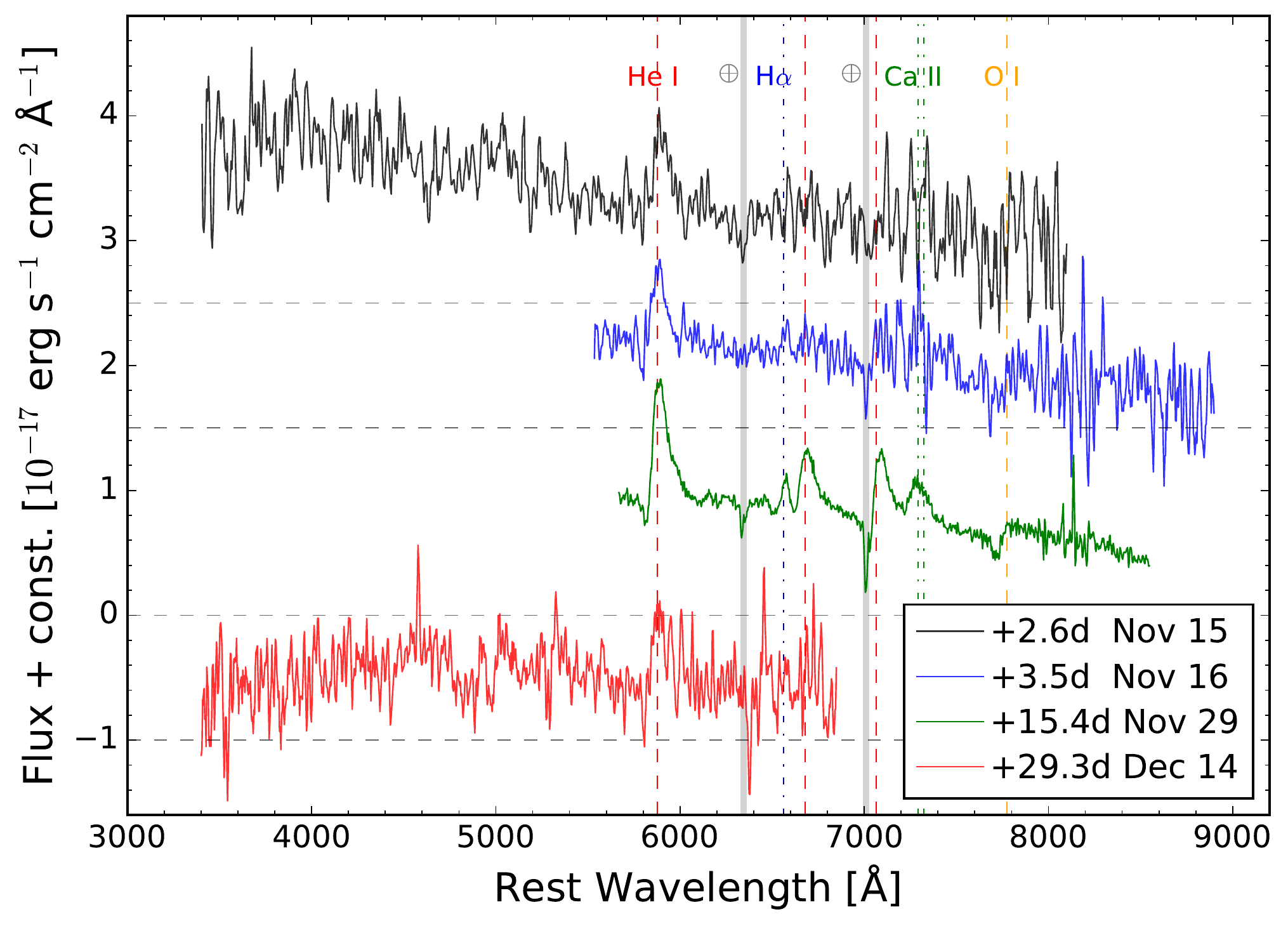}
   \caption{Spectral evolution of OGLE-2014-SN-131. Four spectra were taken with the NTT and the VLT, all showing the prominent \ion{He}{i} feature at 5876~\AA. The zero flux level of each shifted spectrum is indicated with a dashed line. The line identifications from Sect. \ref{sec:spec} are indicated with a red dashed line for \ion{He}{I}, blue dot-dashed line for H$\alpha$, a green dotted line for \ion{Ca}{II}, and a gold dashed line for \ion{O}{I}. The spectra have not been corrected for reddening due to extinction.} 
              \label{specseq}
   \end{figure}

\section{Spectral analysis}
\label{sec:spec}
\begin{figure}
       \centering
       \includegraphics[width=\linewidth]{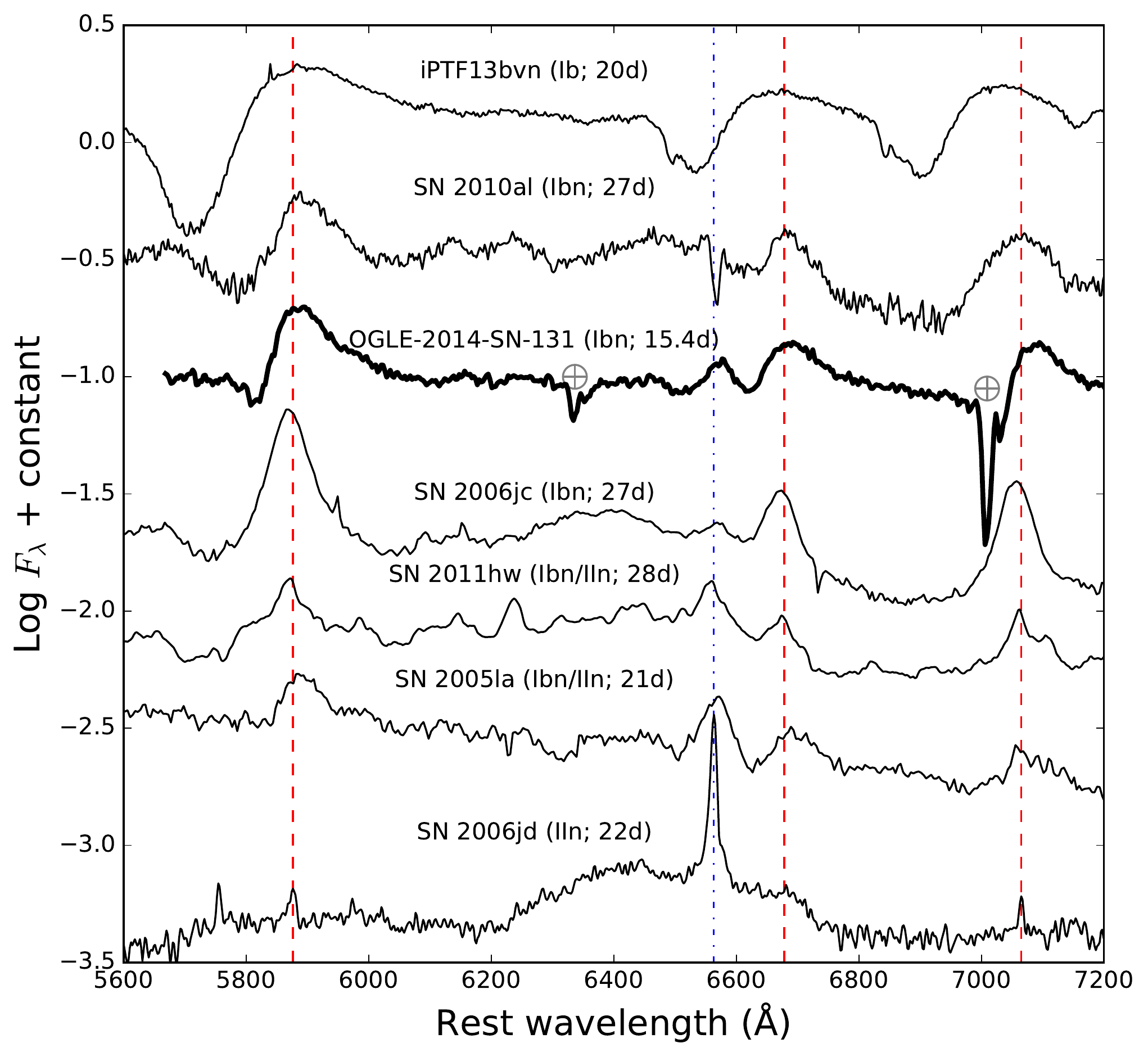}
       \caption{Spectral comparison of OGLE-2014-SN-131 with a sample of SN spectra obtained a few weeks after maximum. The sample includes the Type Ib SN iPTF13bvn \protect\citep{2014A&A...565A.114F}; the Type Ibn SNe 2010al \protect\citep{2015MNRAS.449.1921P} and 2006jc \protect\citep{2008MNRAS.389..113P}; the Type Ibn/IIn SNe 2011hw  and 2005la \protect\citep{2015MNRAS.449.1921P}; and the Type IIn SN 2006jd \protect\citep{2012ApJ...756..173S}. The locations of the prominent \ion{He}{I} lines and that of H$\alpha$ have been marked with red dashed, and a blue dot-dashed line, respectively. This figure is adapted from \protect\citet[their fig. 11]{2015MNRAS.449.1921P}. Phases in parenthesis are rest-frame days since maximum. The main telluric features of our spectrum are marked with crossed-circles.}
       \label{speccomp}
\end{figure}
   
We present our redshift corrected spectral sequence in Fig.~\ref{specseq}.

The VLT spectrum taken on 29 November 2014 prominently displays \ion{He}{i} $\lambda\lambda$5876, 6678, and 7065 emission that allowed us to determine a redshift of $z=0.085\pm 0.002$. Gaussians were fit to the \ion{He}{i} emission peaks at $\lambda$5876 and $\lambda$6678 in order to derive the redshift. There is also a weak emission feature at 6563~\AA\ due to H${\alpha}$ that has a slightly narrower profile than the \ion{He}{i} features.

The most prominent P-Cygni profile feature is the shallow absorption minimum of \ion{He}{i} $\lambda$5876, at V $=3000 \pm100 \text{ km s}^{-1}$. The H${\alpha}$ line also has a shallow blue-ward absorption feature, with a minimum at $2900 \pm100 \text{ km s}^{-1}$, consistent with the \ion{He}{i} $\lambda$5876. Another feature in the spectrum is the emission at 7300~\AA\, which we identify as the blended \ion{Ca}{ii} doublet at $\lambda\lambda$7291, 7323. Finally, there is a feature at 7700 \AA, which could be the absorption part of a P-Cygni line due to \ion{O}{i} $\lambda$7774. If so, this line has a velocity obtained from its minimum of $2900 \pm 800 \text{ km s}^{-1}$, consistent with the \ion{He}{i} $\lambda$5876. 

We used the VLT spectrum to measure the FWHM of the lines.
The \ion{He}{i} $\lambda$5876, where a P-Cygni profile is detected, has a FWHM of $5200 \pm 400 \text{ km s}^{-1}$. The other \ion{He}{i} emission features at 6678 and 7065~\AA,\, where we do not detect a P-Cygni profile, have FWHM velocities of $4700 \pm 400$ and $3300 \pm 400 \text{ km s}^{-1}$, respectively. 
The measurement of the \ion{He}{i} $\lambda$7065 feature appears narrower due to a telluric line that affects its profile. 
The H${\alpha}$ line has a narrower FWHM of $1900 \pm 400 \text{ km s}^{-1}$, while the blended \ion{Ca}{ii} doublet has a FWHM of $3900 \pm 400 \text{ km s}^{-1}$. 

The NTT spectra have low signal-to-noise ratio, and the early spectra are mainly characterized by the presence of the \ion{He}{I} $\lambda$5876 line. Additionally, we can detect H${\alpha}$ in the November 16 spectrum. Aside from these line identifications, no other lines are significantly detected in the NTT spectra.

Based on the presence of strong, but rather narrow (V$_{FWHM} \approx4000 - 5000\text{ km s}^{-1}$), \ion{He}{i} emission lines alongside a weak H${\alpha}$ emission line, we classify OGLE14-131 as a member of the Type Ibn SN class. In Fig.~\ref{speccomp}, the post-maximum VLT spectrum of OGLE14-131 is placed alongside a sequence of post-maximum SN spectra, ranging from SNe~IIn to SNe~Ib. 
The emission features for OGLE14-131 match very well to those of SN~Ibn 2010al \citep{2015MNRAS.449.1921P}. The prominent \ion{He}{i} emission lines are narrower than those typically seen in Type Ib SNe, such as iPTF13bvn, and much stronger than than those sometimes seen in SNe Type IIn, whose spectra are dominated by strong emission lines of hydrogen. 

\begin{figure} 
   \centering
   \includegraphics[width=\linewidth]{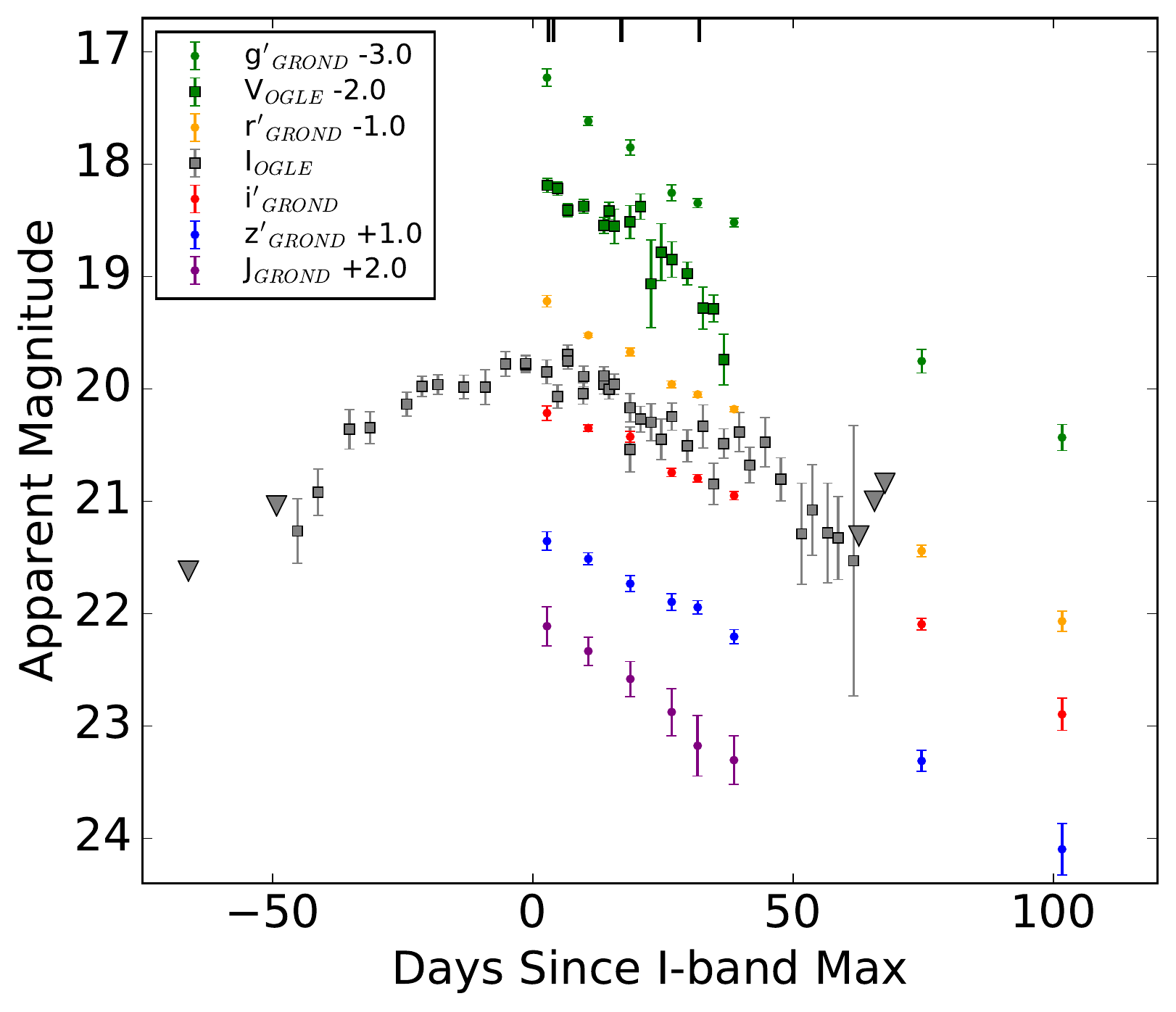}
      \caption{The observed light curves of OGLE-2014-SN-131 in different bands, shifted for easier viewing. The epochs of the obtained spectra have been marked with vertical dashes. $I-$band limits are indicated with gray triangles.}
         \label{lc} 
\end{figure} 


\section{Photometric analysis}
\label{sec:phot}

   \begin{figure*}
   \centering
   \includegraphics[width=15cm]{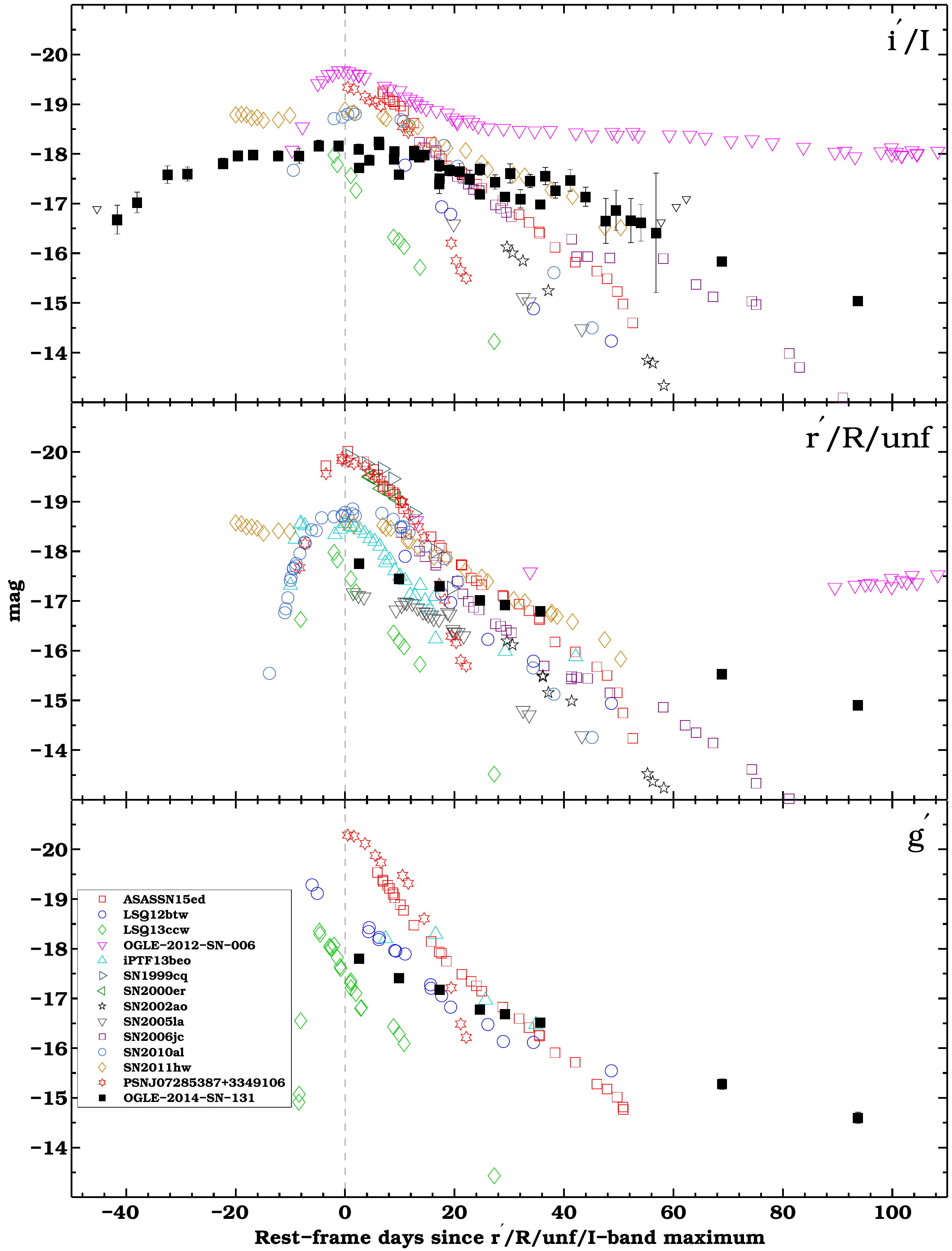}
      \caption{The absolute magnitude light curves of OGLE-2014-SN-131 in the $g',r',i'/I$ filters, compared to those of 
      other SNe~Ibn in the literature \protect\citep[see][and references therein]{2016MNRAS.456..853P}.} 
         \label{compLCIbn} 
   \end{figure*}  

\subsection{The unique light curve of OGLE14-131}
\label{sec:phot1}
While OGLE14-131 is a regular SN~Ibn if we consider its spectrum, its light curve makes it a special event. The unique photometric properties of OGLE14-131 are evident when we compare its light curves to those of other stripped-envelope SNe.

The $I-$band light curve of OGLE14-131 (see Fig.~\ref{lc}) shows a rise time of approximately 45 days in the observer frame (42 days in the rest frame,) during which the light curve rises by $\sim1.5$ mag. Our pre-detection limits are not deep enough to constrain the explosion epoch, so 42 days is the lower limit for the  rise time. The light curve declines by the same amount in the following 52 days (in the rest frame) post peak. The $I-$band light curve peak occurs on JD 2456973.9 $\pm$~1.5 d, as determined by a sextic polynomial fit. We use the JD of $I-$band maximum as phase zero in this paper. The other light curves, $g', r', i', z', J$, observed from the epoch of $I-$band maximum, show a decline similar to that of the $I-$band (see Table~\ref{tab:decline}).

\begin{figure}
    \centering
    \includegraphics[width=\linewidth]{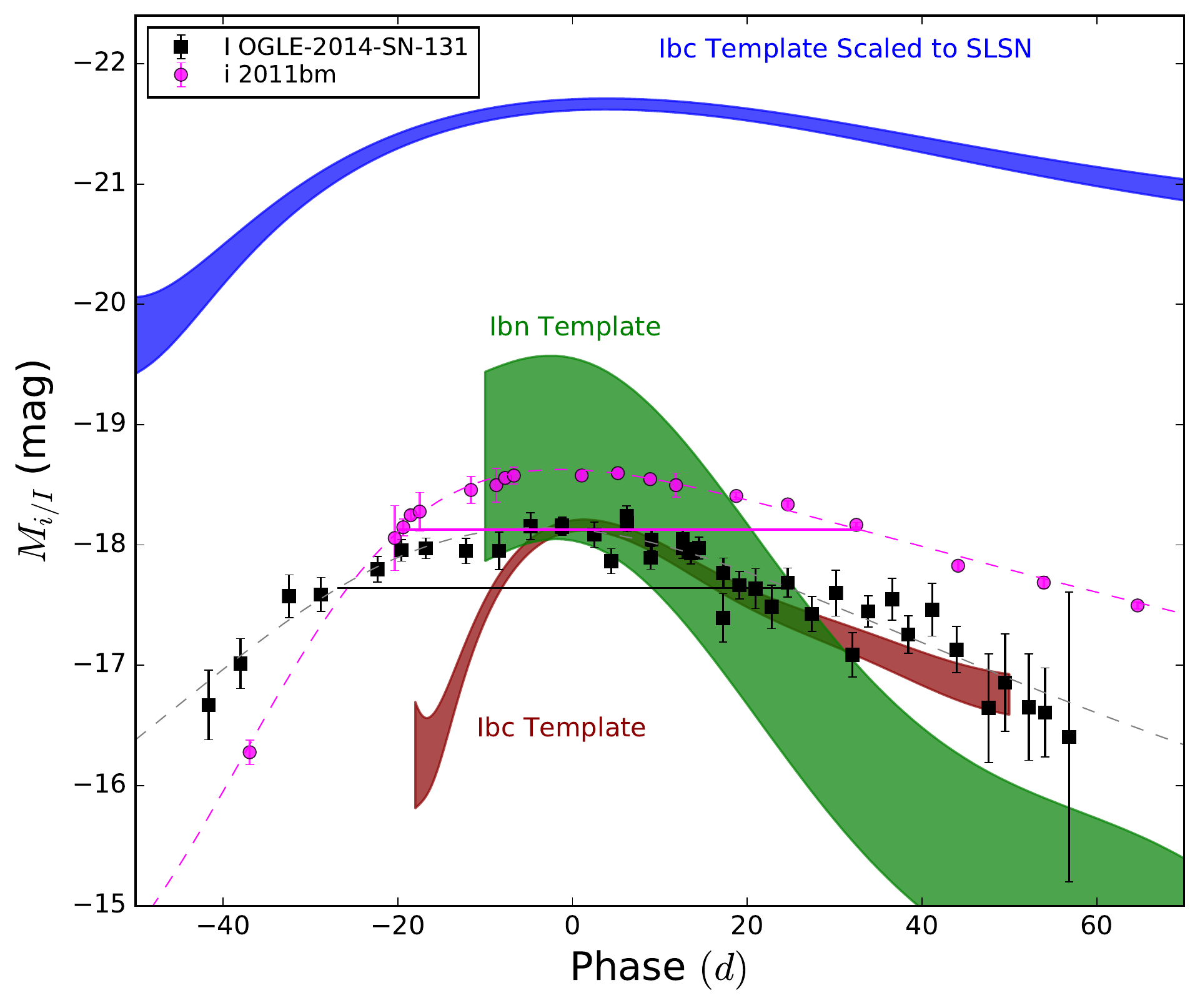}
    \caption{The $I-$band light curve of OGLE-2014-SN-131 alongside Type Ic SN 2011bm \protect\citep{2012ApJ...749L..28V}, compared to template Type Ibn and Type Ibc light curves \protect\citep{2015A&A...574A..60T}, as well as to the SN Ibc template scaled to the SLSN sample \protect\citep{2015MNRAS.452.3869N}. As examples, the phenomenological fits from \protect\citet{2011A&A...534A..43B} used in the broadness comparison for the two SNe have been over-plotted as dashed lines. The broadness parameter plotted in Fig.~\ref{histo_broad} is also indicated with a solid line for both SNe. Both OGLE-2014-SN-131 and SN 2011bm have broader light curves compared to their respective classes. 
    }
    \label{absmagcomp_longrise}
\end{figure}

Using the distance modulus and the extinction correction discussed in Sect.~\ref{sec:disco}, we compute the absolute magnitude light curves of OGLE14-131, which are shown in Fig.~\ref{compLCIbn}. Here we compare the $g'$, $r'$, and $i'/I$-band absolute magnitude light curves of OGLE14-131 with the corresponding light curves of other SNe~Ibn \citep[see][and references therein]{2016MNRAS.456..853P}.    
The $I-$band maximum is found at $-18.16 \pm 0.09$~mag for OGLE14-131,
which is comparable to those found for other SNe~Ibn (see Figs. \ref{compLCIbn} and \ref{absmagcomp_longrise}) and close to the typical value for normal SNe~Ibc \citep{2015A&A...574A..60T}. 
Among SNe~Ibn, OGLE14-131 shows an unprecedentedly long rise-time, as is clearly visible in the top panel of Fig.~\ref{compLCIbn}. SNe~Ibn are usually characterized by fast rise times ($10-15$~days). The decline rate is somewhat slower for OGLE14-131 compared to the other SNe~Ibn, with the exception of OGLE-2012-SN-006.

In Fig.~\ref{absmagcomp_longrise} we also compare the absolute $I-$band light curve of OGLE14-131 to the very slowly evolving light curve of SN~Ic 2011bm. With a well-constrained rise time of $\sim 37-40$~days in the rest frame, SN 2011bm has the slowest photometric evolution of any SN Ibc in the literature \citep{2012ApJ...749L..28V}. Even with a conservative estimate for the explosion epoch as the day of first detection, the rise time of OGLE14-131 is longer than that of SN 2011bm, which is already more than twice as long as an ordinary SN Ibc \citep{2015A&A...574A..60T}. This is shown in Fig.~\ref{absmagcomp_longrise}, where we also plot the average SN Ibn $I-$band light curve (from the data shown in the top panel of Fig.~\ref{compLCIbn})\footnote{The Ibn template is simply the average of SNe Ibn from Fig.~\ref{compLCIbn}, with OGLE-2012-SN-006 excluded due to the unprecedented plateau observed in its light curve.} and the SN~Ibc light curve template \citep[from][]{2015A&A...574A..60T}, as well as a pseudo-template for SLSNe (obtained by scaling the SN~Ibc template according to \citealp{2015MNRAS.452.3869N}). When considering rise-time and broadness, the light curve of OGLE14-131 resembles a super-luminous SN more than a SN~Ibn. It is broader and longer-rising than any other SN of its class. This broadness could offer insights into the progenitor and powering mechanism of OGLE14-131. We study this further in Sect.~\ref{sec:model}. 

The long rise time and the slow decline of OGLE14-131 makes its light curve particularly broad around peak. 
We quantify this peak broadness as the time that the SN light curve spends within half a magnitude of the peak. To compare OGLE14-131 with other light curves in a quantitative fashion, we fit a phenomenological model from \citet{2011A&A...534A..43B} to the $I$/$i$ band for OGLE14-131, the Type Ic SN 2011bm, the Type Ibn and Ibc templates, OGLE14-131's spectroscopic twin Type Ibn SN 2010al, and the rapidly evolving Type Ibn LSQ13ccw ($r$ band, \citealp{2015MNRAS.449.1954P}). In Fig.~\ref{absmagcomp_longrise} we show, as examples, fits to the light curves of OGLE14-131 and SN 2011bm marked by dashed lines, as well as the broadness as a solid line. 

\begin{figure}
 \centering
   \includegraphics[width=\linewidth]{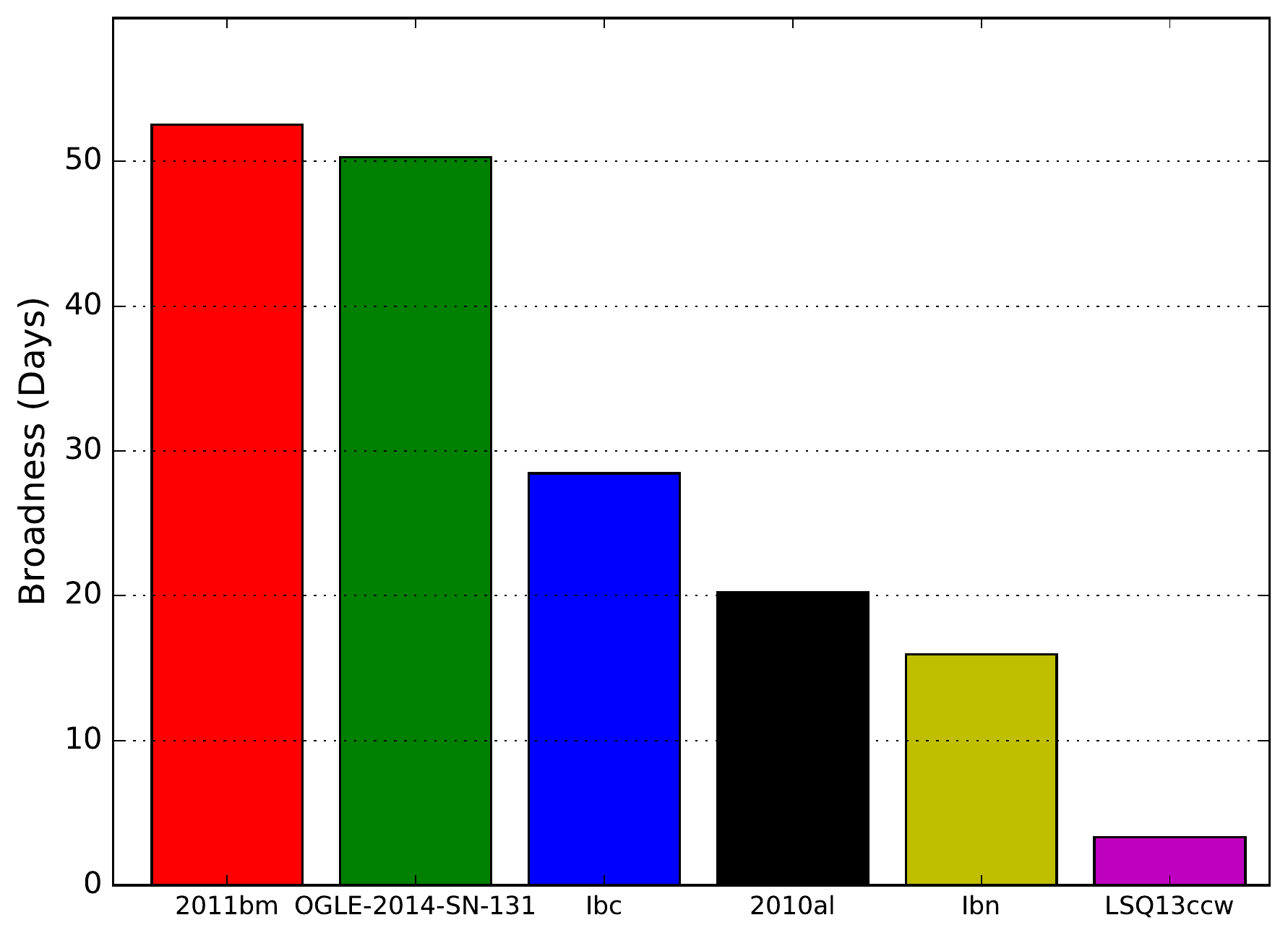}
   \caption{The broadness (time interval when the SN light curve is within 0.5~mag from peak) of the SNe and template light curves discussed in Sect.~\ref{sec:phot1}, and illustrated in Fig.~\ref{absmagcomp_longrise}. (These values characterize the broadness of the light curve around maximum light.) OGLE-2014-SN-131 is broader than normal SNe~Ibc, and than other SNe~Ibn, including its spectroscopic match, SN 2010al. It is instead similar to the broad Type Ic SN 2011bm. We include the rapidly-fading Type Ibn LSQ13ccw to illustrate the diversity of light curve parameters for Type Ibn SNe. Both OGLE-2014-SN-131 and SN 2011bm evolve significantly slower than the typical light curve of their respective classes.}
              \label{histo_broad}
    \end{figure}

This demonstrates that OGLE14-131 is significantly ($\gtrsim \times3$) broader than the typical Type Ibn SN, and similar to SN 2011bm (Fig.~\ref{histo_broad}). 

\subsection{Optical colors of OGLE14-131} 
\begin{figure}
\centering
\includegraphics[width=\linewidth]{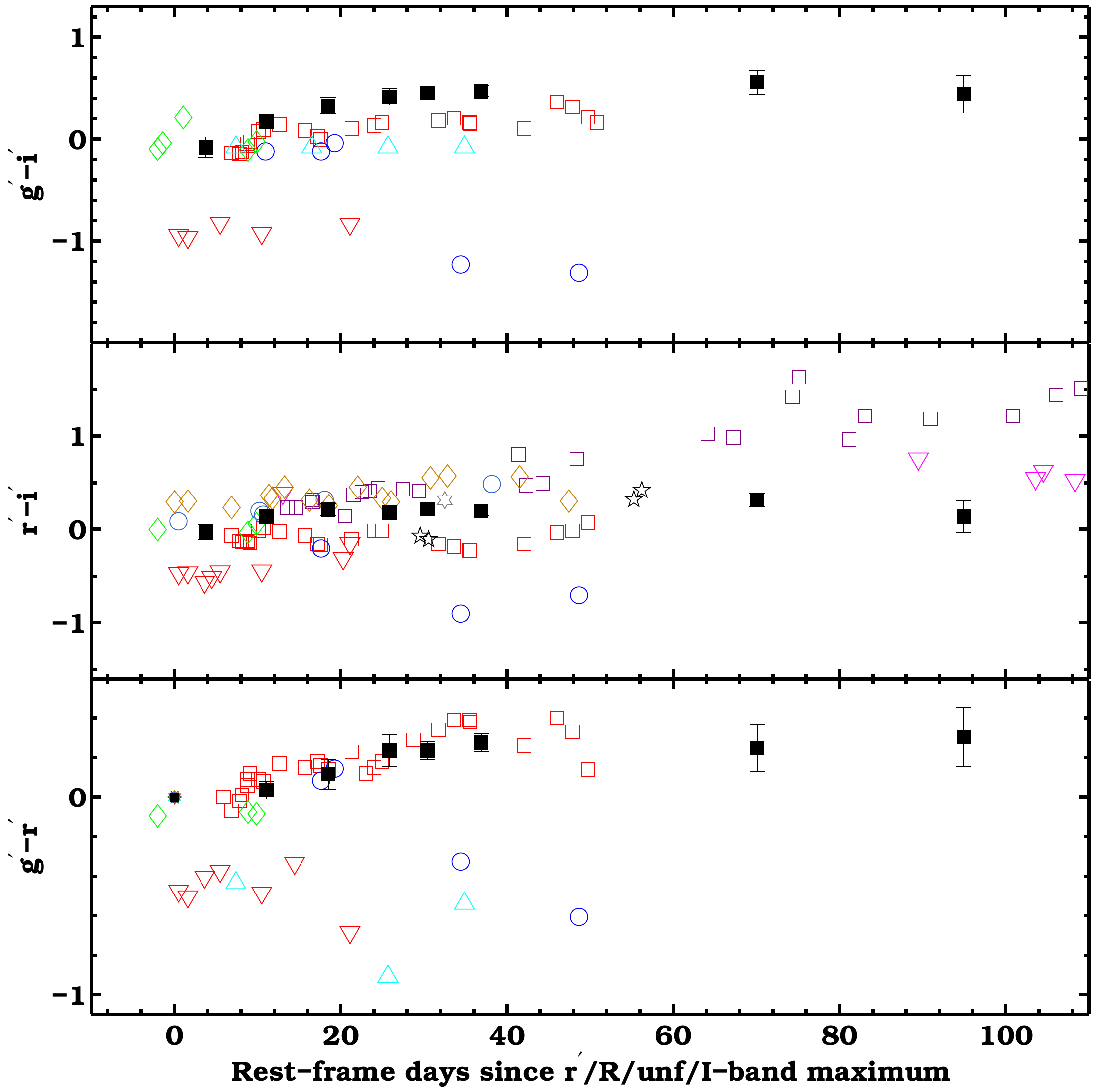}
\caption{Colors of OGLE-2014-SN-131 as compared to those of 
other SNe~Ibn in the literature \protect\citep[see][and references therein]{2016MNRAS.456..853P}. The symbols are the same as in Fig.~\ref{compLCIbn}.}
\label{color}
\end{figure}
The MW extinction corrected optical post-peak colors of OGLE14-131 ($g'-r'$, $g'-i'$, and $r'-i'$) are shown in Fig.~\ref{color} and compared to those of other SNe~Ibn in the literature. The colors for OGLE14-131 gradually evolve blue-ward for $\sim25$~days, before flattening out. Overall, the colors for OGLE14-131 and the other SNe~Ibn appear very similar. The latter exhibit a relatively flat color evolution, or a slow blue-ward trend lasting up to $\sim100$~days.

\subsection{Bolometric properties of OGLE14-131} 
\label{sec:bolo}
\begin{figure}
   \centering
   \includegraphics[width=\linewidth]{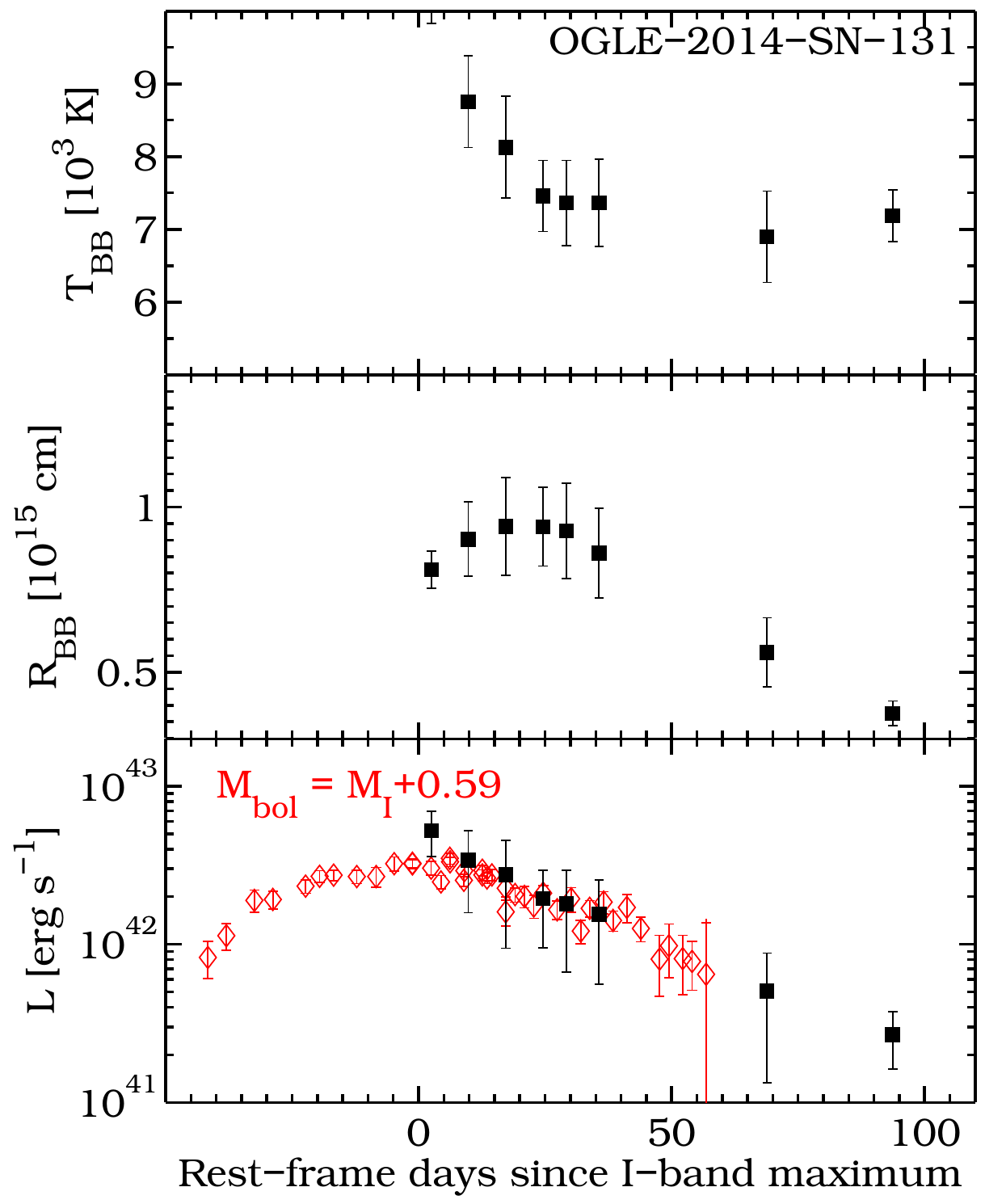}
      \caption{Bolometric properties of OGLE-2014-SN-131 as derived from the BB fit to the SEDs. BB temperature, radius and luminosity are shown by black squares. We only have post-peak SEDs, as before peak OGLE14-131 was only observed in the $I$ band. In the bottom panel we shift the absolute $I-$band magnitude light curve of OGLE14-131 to match its bolometric light curve after peak. We use the combined bolometric plus shifted $I$ band as the final bolometric light curve, in order to also be able to model  its rising part.}
         \label{bolo} 
   \end{figure}

We calculate the bolometric properties of OGLE14-131, namely its bolometric luminosity ($L$), black-body (BB) temperature, and radius. 
To do that, we build a spectral energy distribution (SED) for each epoch of GROND photometry by converting the extinction corrected magnitudes into fluxes. Our SEDs cover the range from $g'$ to $J-$band in effective wavelength. Then we fit the SED (in the rest-frame) with a BB function, obtaining the temperature. We also calculate the luminosity, obtained as the BB integral multiplied by 4$\pi D^2$, where D is the known luminosity distance, in order to estimate the radius. The results are shown in Fig.~\ref{bolo}. The temperature is cooling down in the 20 days post peak, where it remains constant afterwards. The radius reaches its peak at $+$20 days, whereas the luminosity steadily declines. While the SED we constructed was well fit by a BB function, the likely powering scenarios for OGLE14-131 (see Sect.~\ref{sec:discussion},) make it possible that there are non-thermal components contributing to the SED, especially in UV and X-rays.  

If we convert the absolute $I-$band magnitude to luminosity ($L=10^{(-M_I+88.71)/2.5}$~erg~s$^{-1}$), we can see that, after a bolometric correction of $+$0.59 mag, it is very similar to the computed bolometric luminosity at the epochs post peak (see  the red diamonds in Fig.~\ref{bolo} as compared to the black squares). We obtain this bolometric correction by fitting the bolometric epochs to the $I-$band light curve where the two overlap, and assume this shift to be the bolometric correction for the $I-$band light curve that is constant in time. We then merge the two bolometric light curves, in order to also cover the rising part. We model this pseudo-bolometric light curve in Sect.~\ref{sec:model}. 

The validity of assuming a bolometric correction that is constant in time depends on the flatness of the color evolution of Type Ibn SNe. While there is no such thing as typical color changes for SNe Ibn, the post peak colors of OGLE14-131 are in good agreement with others from the literature, (we refer to Fig.~\ref{color}). The colors before maximum are harder to find. For the well studied SN 2010al, the colors are relatively flat before peak, within a range of 0.25 mag. This is also the case for PS1-12sk \citep{2013ApJ...769...39S}, whose pre-peak colors are flat within 0.1 mag. However, OGLE14-131 has an unprecedented early light curve, and we do not know its early color evolution. This is a possible source of uncertainty for the models we use in the following section. 

\begin{table}
\centering
    \begin{tabular}{lll}
    \hline
    \noalign{\smallskip}
    Band & Peak & Slope \\
    \noalign{\smallskip}
    \hline
    \noalign{\smallskip}
    \_ & (mag) &  (mag d$^{-1}$) \\
    \noalign{\smallskip}
    \hline
    \noalign{\smallskip}
    I-OGLE      & 19.77 $\pm$ 0.075 & 0.025 \\ 
    V-OGLE      & 20.19-20.05  & 0.034 \\
    g$'$-GROND  & 20.23-20.23$^\text{a}$ &  0.032 \\
    r$'$-GROND  & 20.22-20.15 &  0.029 \\
    i$'$-GROND  & 20.21-20.00 &  0.027 \\
    z$'$-GROND  & 20.35-20.18 &  0.028 \\
    J-GROND     & 20.11-19.98 &  0.035 \\
    \noalign{\smallskip}
    \hline
    \end{tabular}
    \caption{Peak magnitude and decline rate of each photometric band. Decline rates were obtained by fitting a first degree polynomial to all epochs post peak for a given filter. Peak magnitude for all bands except for the $I$ band are given as ranges, where the upper limits have been extrapolated from the decline rate fit at the JD of $I-$band maximum, since they lack data at peak. We consider the fit values to be upper limits since a concave-down light curve would have a dimmer peak than this fit indicates. The lower limits are simply the magnitude of first detection immediately after peak. The peak magnitude values have not been extinction corrected. \protect\\ \protect$^\text{a}$ The peak magnitude for $g'$ band was estimated by excluding the last epoch of photometry due to it slightly diverging from a simple linear fit. The decline rate is still from a fit to all epochs.}
    
    \label{tab:decline}
\end{table}

\section{Modeling}
\label{sec:model}

\begin{figure}
\centering
\includegraphics[width=\linewidth]{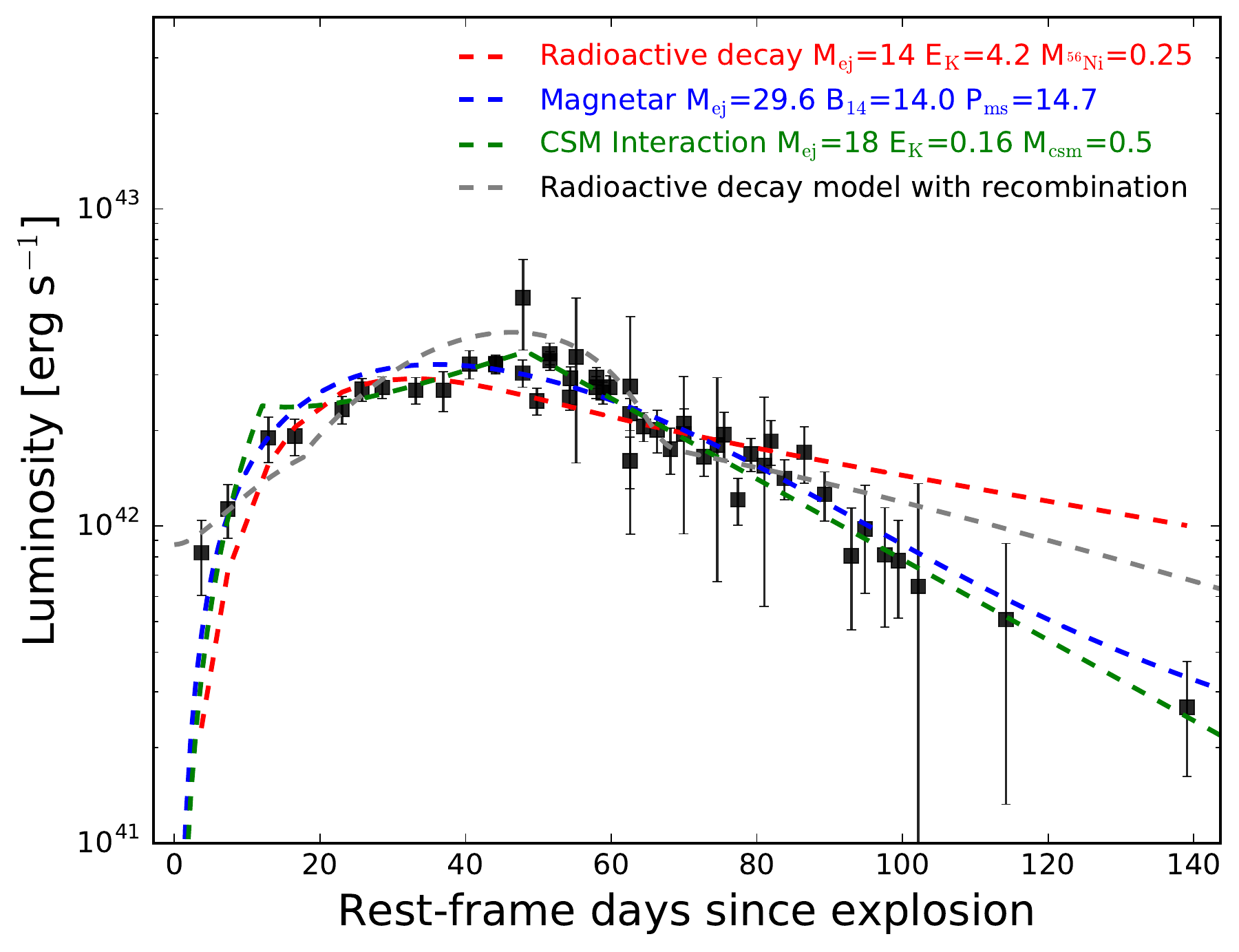}
\caption{Best fits to the bolometric light curve of OGLE-2014-SN-131 for three different models. The dashed red line shows the simple Arnett (radioactive decay) model, the dashed blue line shows the best magnetar model, the dashed green line indicates the best CSM interaction model, and finally, the dashed gray line is an alternative radioactive decay model with recombination. The radioactive decay models cannot reproduce the relatively fast decline of the light curve. 
} 
\label{model} 
\end{figure} 

The light curves of SNe~Ibn have been explained by invoking radioactive decay
or circumstellar interaction as the powering mechanisms. The energy source of the emission from long-rising SE~SNe, such as SN 2011bm and SN 2005bf, has been proposed to be $^{56}$Ni radioactive decay and the spin-down of a magnetar, respectively. OGLE14-131 could be powered by one, or a combination, of these mechanisms. Though there are clear signs of CSM interaction (the spectra display narrow emission lines similar to those in SNe~IIn and other SNe~Ibn; see Fig.~\ref{specseq}), the dominant powering source could be something else. In the following, we discuss these different mechanisms for OGLE14-131 by fitting simple analytical models to the pseudo-bolometric light curve. Our simple modeling is suitable for estimating the relevant explosion parameters for a given powering scenario, which is the aim of this section. For this discussion, we assume the last non-detection epoch as the explosion date. 

\subsection{Radioactive decay}
Assuming that radioactive decay of $^{56}$Ni is the dominant energy source, we can fit a simple \citet{1982ApJ...253..785A} (radioactive decay) model \citep[see e.g., ][]{2013MNRAS.434.1098C,2015A&A...574A..60T} to the bolometric light curve of OGLE14-131. We obtain an ejecta mass M$_\text{ej}$ $=$ 14 M$_\odot$, a $^{56}$Ni mass M$_{^{56}\text{Ni}}$ $=$ 0.25 M$_\odot$, and an explosion energy E$_\text{K}=4.2\times10^{51}$~erg. We plot our best fit with a red-dashed line in Fig. \ref{model}. As can be seen in the figure, it is difficult to fit the rise, peak, and decline simultaneously using a simple Arnett fit. Such an amount of $^{56}$Ni combined with the large ejecta mass should keep the light curve brighter for longer post peak, which does not match our observations. This suggests that the sole powering source is not radioactive decay. 

In the Arnett model, we assumed $\kappa=0.07$~cm$^2$~g$^{-1}$ (which is well suited for electron scattering in H-free material), and an ejecta expansion velocity V~$=7000$~km~s$^{-1}$. As we cannot see the spectrum of the ejecta of OGLE14-131, which is masked by the slowly moving CSM, we resort to using the above-mentioned value for the ejecta velocity, since it was measured for the SN Ibn/Ib ASASSN15ed \citep{2015MNRAS.453.3649P}. If the expansion velocity used in the Arnett model was higher, a greater ejecta mass would be needed to get a good fit, and vice versa for a lower expansion velocity. In the Arnett model, we also use $E/M=(3/10)V^2$ from the assumption of constant density.

The mismatch between the long rise and the fast decline compared to the Arnett model could possibly be resolved if we imagine that after peak the gamma-rays powering the light curve are not fully trapped. To investigate this possibility, we used a scaling-relation for gamma-ray escape \citep{1997ApJ...491..375C}, but also included a treatment for the 3.5 percent of kinetic energy released by positrons \citep[e.g.,][]{1998A&A...337..207S}, with the M$_\text{ej}$ and E$_\text{K}$ parameters from our Arnett model. Using this method, assuming that gamma-ray escape scales as $\propto 1-0.965 \exp(-\tau)$, we find that even for the final epoch, the gamma-rays are 99.3\% trapped, and thus our assumption of full-trapping in the Arnett model is valid. Since $\tau \propto (M_\text{ej}^2 / E_K) t^{-2}$, we also calculated a one sigma upper limit of M$_\text{ej}$ $\lesssim 4.2$ M$_\odot$ to explain the final photometric epoch by gamma-ray escape, as compared to our Arnett model with full trapping, when keeping the kinetic energy fixed. However, $4.2$ M$_\odot$ of ejecta mass is too low to explain the broadness of the lightcurve, and therefore 
we disfavor this scenario. 

While the mass of $^{56}$Ni we obtain is similar to that found in other SNe Ibc \citep{2016MNRAS.457..328L}, the ratio of the $^{56}$Ni mass to the ejecta mass, which is approximately $~2\%$, is relatively low. CC SNe and SE SNe generally show a correlation between the synthesized  $^{56}$Ni mass and the ejecta mass \citep{2016MNRAS.457..328L}. For an ejecta mass $>$10~M$_{\sun}$, a SE~SN  powered by radioactive decay typically produces $\gtrsim$~0.7~M$_{\sun}$ of $^{56}$Ni, as in the case of SN~2011bm \citep{2012ApJ...749L..28V}. 
This is not the case for OGLE14-131, whose putative $^{56}$Ni mass is substantially lower.

Based on all of the above, we do not favor the simple Arnett model. We also briefly investigated whether composite diffusion models \citep{1992AZh....69..497I,1989ApJ...340..396A}, which include the effect of helium recombination in the SN ejecta, could reproduce the light curve of OGLE14-131. We implemented the semi-analytical model from \cite{1992AZh....69..497I}, and obtain an acceptable fit to the early light curve using M$_\text{ej}$ $=$ 6 M$_\odot$, M$_{^{56}\text{Ni}}$ $=$ 0.23 M$_\odot$, E$_\text{K}=1.4\times10^{51}$~erg, as well as an extended stellar radius of 100 R$_\odot$. This alternative model is plotted with gray dashed lines in Fig. \ref{model}. While the hybrid diffusion model can fit the early part of the broad light curve of OGLE14-131, since recombination mainly affects the early light curve, the radioactive decay scenario still struggles to reproduce the relatively fast late decline.
 
\subsection{Magnetar}

We can also try to fit the bolometric light curve with a magnetar model, first proposed and applied to supernovae by \cite{2010ApJ...717..245K,2010ApJ...719L.204W}. Assuming constant opacity $\kappa~=~$0.07~cm$^2$~g$^{-1}$, V = 7000~km~s$^{-1}$, and uniform density of the ejecta (as in the Arnett model), we fit the bolometric light curve to estimate the ejecta mass, the strength of the magnetic field ($B_{14}$) in units of 10$^{14}$~G, and the spin period of the magnetar in ms ($P_{ms}$), as outlined by \cite{2013ApJ...770..128I}. We obtain a large ejecta mass of 30 $\pm$ 6 ~M$_{\odot}$, $B_{14}$\footnote{This $B$ field is similar to that estimated for SN~2001gh \citep{2015MNRAS.451.3151E}.}$~=~$14 $\pm$ 1, and P$_{ms}~=$ 15 $\pm$ 2. The best fit is shown with a blue-dashed line in Fig.~\ref{model}.

Whereas we cannot rule out the magnetar scenario, we do not need to invoke it for additional luminosity, and we do not pursue this idea further. Since we do see evidence of CSM interaction in the spectra, the simplest scenario is probably that the light curve is also mainly powered by CSM interaction, which we discuss immediately below.

\subsection{Circumstellar interaction}
\label{sec:csm}

The spectra and a relatively fast decline of the light curve can be more easily explained in the context of CSM interaction. The presence of narrow emission lines in the spectrum of our SN~Ibn suggests that some CSM-interaction is certainly occurring. Here we investigate whether this mechanism could be the dominant source of energy for OGLE14-131. We adopt the semi-analytic light curve model by \cite{2013ApJ...773...76C}, which models the effects of forward and reverse shock energy deposition, taking into account diffusion, through an optically-thick CSM assuming that the typical shock crossing timescale is larger than the effective diffusion timescale.  

This model includes several parameters characterizing the progenitor star and the explosion mechanism (ejecta mass, explosion energy, radius, inner and outer density structure) and the CSM (density, density structure, opacity). Obviously, the large number of parameters makes it easier to fit a light curve. In particular, we can choose several sets of parameters to reproduce both the rise and the linear (in magnitude) decline. 
In Fig.~\ref{model} we show one example of a good CSM-interaction model fit (green dashed line; see appendix for model parameters). The CSM model can reproduce both the parabolic early light curve, and the linear decline part of the light curve after peak rather well. 

The dense CSM ($\sim 10^{-10}$ g cm$^{-3}$) in the models can be produced by an eruption or by a steady-wind prior to collapse. While the exact formation scenario is not known, in accordance with previous modeling work for Type Ibn SNe, (e.g., \citealp{2009MNRAS.400..866C,2015MNRAS.451.3151E,Moriya2016},) we assume a constant-density CSM shell in our model. This putative shell of CSM may have been ejected prior to explosion in an eruptive mass-loss episode, such as the pre-explosion outburst observed for SN 2006jc. We also tried using a model with a CSM formed by steady-state mass-loss, but the high mass-loss rate required to form the dense CSM for this model was incompatible with a steady wind scenario.

\begin{figure}
\centering
\includegraphics[width=\linewidth]{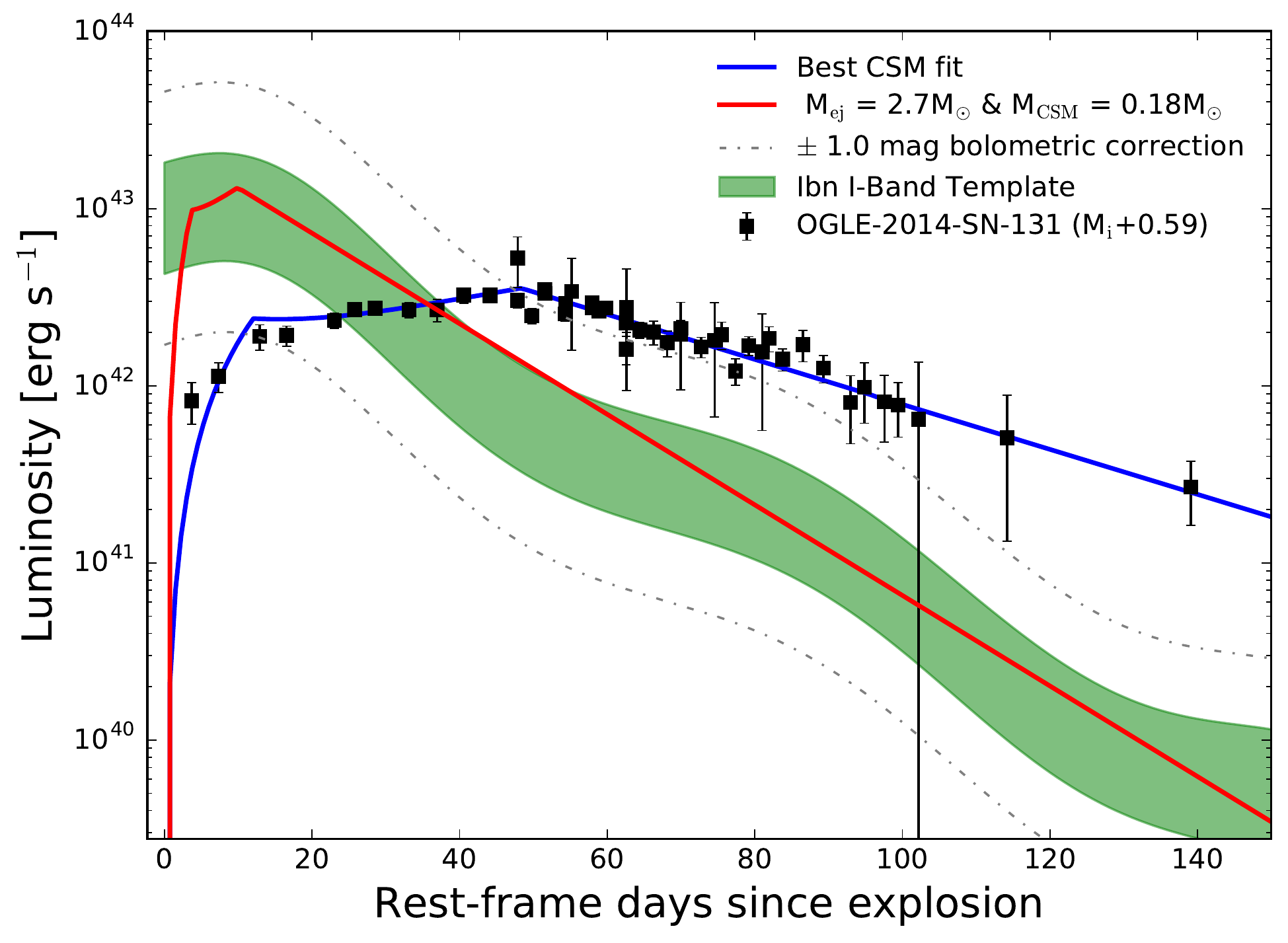}
\caption{CSM interaction models compared to the light curve of OGLE-2014-SN-131 and the $I$~band SN Ibn template. The only difference between the two models is that the latter (red line) had its ejecta mass and CSM mass reduced by 85\% and 65\%, respectively. No other parameter has been changed. Since we do not know the average $I$-band bolometric correction for a SN Ibn, we also plot the 1-sigma $I$-band template with $\pm $ 1.0 mag bolometric correction. The model would still be consistent with the template after we apply this bolometric correction, if we allow the explosion energy to change by $\pm 30$\% (see Sect.~\ref{sec:csm}).}
\label{CSM} 
\end{figure}

In Fig.~\ref{CSM}, we present again the best CSM model for OGLE14-131 (blue line), as well as another model (red line) that better fits the fast rise of normal SNe~Ibn and their early declining phase. For illustration, we also plot the SN Ibn template. The two models have the same parameters controlling the early rising part of the light-curve, except that the fast rising one is built by reducing both the ejecta mass and the CSM mass by 85\% and 65\%, respectively. The reduction in expelled mass has the effect of both shortening the rise time, and increasing the peak luminosity slightly. 
 
However, these are not the only parameters we can change to fit the general shape of a SN Ibn light curve. For example, increasing the explosion energy has the effect of increasing the peak luminosity, without a significant effect on the rise-time \citep{2015MNRAS.452.3869N}. We can also modify the progenitor radius, explosion energy, CSM mass, and the parameters responsible for the CSM density profile together to achieve a similar result. We did not attempt to establish a grid of possible parameters. However, Fig. \ref{CSM} illustrates that by simply varying the CSM and ejecta masses within a physically plausible range, we can reproduce the observed light curve quite comfortably.

Though we do not fit hybrid models due to the large number of parameters involved, (a hybrid model would include all of the parameters of each individual powering scenario,) these multiple powering scenario models, such as a combination of $^{56}$Ni decay and CSM interaction \citep{2013ApJ...773...76C}, could potentially explain the light curve of OGLE14-131, as well. 

\section{Discussion}
\label{sec:discussion}
Given the spectral properties of OGLE14-131, the unprecedented long rise-time, and the broadness of its light curve, as well as the model fits to the bolometric light curve: 
1) we disfavor the radioactive decay scenario, which cannot reproduce the fast decline and the long rise simultaneously, while still fitting the peak luminosity; 
2) we cannot rule out the magnetar scenario, whose model fits well to the bolometric light curve;
3) we favor the CSM interaction scenario, which can potentially explain all the observed properties. 

In the CSM interaction scenario, we can also attempt to explain why OGLE14-131 rises over a longer timescale than usual for SNe~Ibn. In terms of the progenitor star, we make the zeroth order assumption that larger ejecta and CSM mass should be produced by a star that is more massive than the typical SN~Ibn progenitor. Then, if SNe Ibn have a rather uniform set of progenitors \citep{griffin2016}, and we therefore can use similar CSM interaction model parameters, we show in Fig. \ref{CSM} that the primary difference between a typical SN Ibn and the unusual light curve of OGLE14-131 could be due to the mass of the progenitor. Hence, it is reasonable to conclude that what makes OGLE14-131 special in the context of SNe~Ibn is the nature of its progenitor star, which was more massive than the usual SN~Ibn progenitors. 

The large ejecta mass inferred for OGLE14-131 has implications for the progenitor evolution. In the single-star Binary Population and Spectral Synthesis code \citep[BPASS]{2009MNRAS.400.1019E}, models at Large Magellanic Cloud metallicity ($\sim 1/2$ solar), are able to lose their H envelope and form a large enough CO core-mass to explain our ejecta mass estimate, at M$_\text{ZAMS}$ between 40 to 60 M$_\odot$. OGLE14-131 evolved in a low-metallicity environment (see Sect.~\ref{sec:disco}), and therefore, it might have been produced by a massive single star, according to these models. Another possibility is that the progenitor was stripped of its H envelope by a binary companion \citep{2008MNRAS.384.1109E}. A massive progenitor near 60 M$_\odot$ in a low-metallicity environment with rapid rotation could have an eruptive mass-loss history via the pulsational pair instability (PPI; \citealp{2016arXiv160808939W}).

From the light curve fitting (Sect. \ref{sec:csm}), we also favor the presence of a CSM shell around the SN progenitor instead of a wind-like profile. While the exact eruptive mechanism responsible is not known, gravity-wave induced mass-loss \citep{2012MNRAS.423L..92Q}, luminous blue variable (LBV) type mass-loss, or a PPI \citep{2007Natur.450..390W,2016arXiv160808939W} could have formed the CSM shell. The PPI mechanism is expected to naturally result in CSM shells \citep{2007Natur.450..390W}.

The light curve of OGLE14-131 is broader and longer-rising than any other SN of its type. This is true also for SN 2011bm, which was likely the explosion of a very massive (M$_\text{ZAMS}$ $\geq 30$~{M}$_\odot$, and M$_\text{ej} \sim 7-17$ M$_\odot$) star \citep{2012ApJ...749L..28V}. The similarity in relative light curve broadness and rise-time of the two SNe as compared to their respective types could indicate that OGLE14-131 also has an unusually massive progenitor, if the light curve is nickel powered. Even in the CSM powering scenario (Sect.~\ref{sec:csm}), we show that the broad light curve of OGLE14-131 compared to its class can be due to an unusually massive progenitor.

The relatively fast ($\sim$3000 km s$^{-1}$) CSM velocity that can be deduced from the absorption minima in the spectrum is consistent with the wind velocities of very massive WR stars, especially early-type WC and WO stars \citep{2007ARA&A..45..177C}. A very large $M_{ZAMS}$ star that evolved into a WR star more massive than usual likely produced OGLE14-131.

\citet{griffin2016} showed that SNe~Ibn exhibit remarkable homogeneity in their light curves for circumstellar interaction powered SNe. Since slowly-evolving light curves should be more readily observed, there is no observational bias against the rarity of broad light curves in SNe Ibn. However, no other SN Ibn, with as long a rise-time as OGLE14-131, has been published in the literature. In fact, \citet{griffin2016} showed that SNe~Ibn exhibit remarkable homogeneity in their fast-rising light curves. Therefore, the progenitor system and/or the CSM configuration around OGLE14-131 must be unusual compared to an ordinary SN~Ibn. If, as we argue, OGLE14-131 has an unusually massive progenitor for a SN~Ibn, then it is natural to expect a low number of OGLE14-131-like events, given that such massive stars should be relatively rare.
 
\section{Conclusions}
\label{sec:conclusion}

Spectroscopy and multi-band photometry of OGLE14-131 have been presented. Identification of prominent \ion{He}{I} emission lines classifies this SN as a Type Ibn. While, spectroscopically, OGLE14-131 is identified as a normal SN Ibn, when compared to other Type Ibn SNe, its broad light curve shows the longest rise-time ever observed for this SN class. From the post-peak multi-band photometry, a pseudo-bolometric light curve was computed, and combined with the $I-$band light curve (after applying a bolometric correction), to also cover the rising part. Analytical light curve models fitted to this pseudo-bolometric light curve disfavor a radioactive decay powered scenario. Instead, the CSM interaction scenario, which is also supported by the narrow emission line spectra of OGLE14-131, is preferred. 

CSM model fits to the long rise-time require higher ejecta and CSM masses for the progenitor of OGLE14-131 compared to typical SN Ibn progenitors. Given the composition of the CSM inferred from the spectra, and the long rise to peak, the most likely progenitor scenario for OGLE14-131 is a massive WR star. The relatively high wind velocity of $\sim3000$ km s$^{-1}$ means that it is likely a WC or WO star, which can show faster winds. 

\begin{acknowledgements}  
The authors would like to acknowledge S. Klose for his contribution from GROND and A. Pastorello for his insightful comments. We are also grateful to J. Bolmer, C. Delvaux, T. Schweyer, M. Tanga, and K. Varela (all GROND team, MPE Garching) for the support during the GROND observations. We would like to thank the referee for timely and helpful suggestions during the peer-review process. We thank A. Nyholm and M. Ergon for their informative discussions and suggestions during many weekly meetings. L. Wyrzykowski acknowledges the National Science Centre’s grant no. 2015/17/B/ST9/03167. S. Schmidl acknowledges support by DFG grant Kl 766/16-1 and the Th\"uringer Ministerium f\"ur Bildung, Wissenschaft und Kultur under FKZ 12010-514. S.~J. Smartt acknowledges funding from the European Research Council under the European Union's Seventh Framework Programme (FP7/2007-2013)/ERC Grant agreement n$^{\rm o}$ [291222] and STFC grants ST/I001123/1 and ST/L000709/1. M. Sullivan acknowledges support from EU/FP7-ERC grant 615929 and STFC grant ST/L000679/1.    

We gratefully acknowledge support from the Knut and Alice Wallenberg Foundation. This work is based on observations collected at the European Organisation for Astronomical Research in the Southern Hemisphere, Chile as part of PESSTO, (the Public ESO Spectroscopic Survey for Transient Objects Survey) ESO programme 191.D-0935. This research has made use of the NASA/IPAC Extragalactic Database (NED) which is operated by the Jet Propulsion Laboratory, California Institute of Technology, under contract with the National Aeronautics and Space Administration. The Oskar Klein Centre is funded by the Swedish Research Council. The OGLE project has received funding from the National Science Centre, Poland, grant MAESTRO 2014/14/A/ST9/00121 to AU. This work has been supported by the Polish Ministry of Science and Higher Education through the program “Ideas Plus” award No. IdP2002 000162 to IS.  Part of the funding for GROND (both hardware as well as personnel) was generously granted from the Leibniz-Prize to Prof. G. Hasinger (DFG grant HA 1850/28-1). This work was partly supported by the European Union FP7 programme through ERC grant number 320360. This work is based on observations collected under ESO DDT proposal no 294.D-5011.

      \end{acknowledgements}
      
\bibliographystyle{aa}
\bibliography{ogle131.bib}

\begin{appendix}
\section{CSM model parameters}
\label{csmpars}
Since the CSM model has many parameters, we did not attempt to optimize the fit to our light curve. Instead, we used our intuition of the physically based parameters to obtain a visually acceptable fit. For simplicity, we assumed a purely CSM powered scenario with no nickel. 
We used the following values for the nature of the CSM: Mass of the CSM, M$_\text{CSM}$ = 0.5 M$_\odot$; the power-law exponent for the CSM density profile, s = 0; the density of the CSM shell immediately outside the stellar envelope, $\rho_\text{CSM}$ = $0.55 \times 10^{-10}$ g cm$^{-3}$. 
We used the following values characterizing the nature of the progenitor: the inner density profile of the ejecta, $\delta=1$; the explosion energy E$_\text{SN}$ = 0.155 $\times 10^{51}$ ergs; the radius of the progenitor, R$_\text{p}$ = 1 R$_\odot$; the ejecta mass, M$_\text{ej}$ = 18 M$_\odot$; the power-law exponent of the outer ejecta component, n = 12. 
To set the diffusion timescale of the optically thick CSM, we also used $\kappa=0.2$ cm$^2$ g$^{-1}$ for the opacity and a constant of $\beta = 13.8$.

For the model better fitting a typical SN Type Ibn, we did not change any of the parameters above except for the ejecta and CSM masses, which we multiplied by a constant of $0.15$ and $0.35$, respectively. The ejecta and CSM masses were changed to better fit the rapid rise and slightly-higher peak luminosity of a typical SN~Ibn.

\section{Photometry tables}

\clearpage

\begin{table}
 \begin{minipage}{\linewidth} 
    \begin{tabular}{l l l l l l}
            \hline
            \noalign{\smallskip}
            MJD & I-Mag & I-Error & MJD & V-Mag & V-Error \\
            \noalign{\smallskip}
           \hline
            \noalign{\smallskip}
        56907.232 &  \textgreater21.62   &   \_     & \_     & \_  & \_  \\
        56924.169 &  \textgreater21.04   &   \_     &  \_    & \_  & \_  \\
        56928.197 &  21.26    &   0.29 & \_  & \_  & \_  \\
        56932.130 &  20.92    &   0.21 & \_  & \_  & \_  \\
        56938.167 &  20.36    &   0.18 & \_  & \_  & \_  \\
        56942.125 &  20.34    &   0.14 & \_  & \_  & \_  \\
        56949.152 &  20.14    &   0.11 & \_  & \_  & \_  \\
        56952.129 &  19.98    &   0.09 & \_  & \_  & \_  \\
        56955.145 &  19.96    &   0.09 & \_  & \_  & \_  \\
        56960.132 &  19.98    &   0.10 & \_  & \_  & \_  \\
        56964.265 &  19.98    &   0.16 & \_  & \_  & \_  \\
        56968.162 &  19.78    &   0.11 & \_  & \_  & \_  \\
        56972.054 &  19.78    &   0.07 & \_  & \_  & \_  \\
        56972.054 &  19.77    &   0.07 & \_  & \_  & \_  \\
        56976.080 &  19.85    &   0.11 &  56976.171 &  20.19    &     0.06  \\  
        56978.175 &  20.07    &   0.10 &  56978.172 &  20.22    &     0.06  \\  
        56980.096 &  19.70    &   0.09 &  56980.130 &  20.41    &     0.06  \\
        56980.132 &  19.75    &   0.07 & \_  & \_  & \_                                          \\
        56983.066 &  20.04    &   0.10 &  56983.180 &  20.37    &     0.06  \\
        56983.182 &  19.89    &   0.09 & \_  & \_  & \_                                         \\  
        56987.044 &  19.96    &   0.09 &  56987.042 &  20.54    &     0.07  \\
        56987.064 &  19.88    &   0.08 & \_  & \_  & \_                                         \\ 
        56988.059 &  20.00    &   0.09 &  56988.057 &  20.41    &     0.08  \\
        56989.073 &  19.96    &   0.09 &  56989.071 &  20.55    &     0.15  \\
        56992.060 &  20.54    &   0.20 &  56992.058 &  20.51    &     0.15  \\
        56992.080 &  20.17    &   0.13 & \_  & \_  &  \_                                        \\
        56994.086 &  20.27    &   0.11 &  56994.084 &  20.38    &     0.11\\
        56996.092 &  20.30    &   0.17 &  56996.101 &  21.06    &     0.39\\
        56998.089 &  20.45    &   0.18 &  56998.087 &  20.78    &     0.25\\
        57000.099 &  20.25    &   0.12 &  57000.108 &  20.85    &     0.16\\
        57003.081 &  20.51    &   0.14 &  57003.079 &  20.97    &     0.10 \\
        57006.089 &  20.33    &   0.19 &  57006.070 &  21.28    &     0.19 \\
        57008.133 &  20.85    &   0.18 &  57008.131 &  21.29    &     0.12 \\
        57010.074 &  20.49    &   0.13 &  57010.097 &  21.74    &     0.23 \\
        57013.062 &  20.39    &   0.17 & \_  & \_  &  \_  \\
        57015.069 &  20.68    &   0.16 & \_  & \_  &  \_  \\
        57018.030 &  20.47    &   0.22 & \_  & \_  &  \_  \\
        57021.048 &  20.80    &   0.19 & \_  & \_  &  \_  \\
        57025.037 &  21.29    &   0.45 & \_  & \_  &  \_  \\
        57027.077 &  21.08    &   0.41 & \_  & \_  &  \_  \\
        57030.032 &  21.28    &   0.44 & \_  & \_  &  \_  \\
        57032.044 &  21.33    &   0.37 & \_  & \_  &  \_  \\
        57035.030 &  21.53    &   1.20 & \_  & \_  &  \_  \\
        57036.031 &  \textgreater21.31    &   \_ & \_  & \_  & \_  \\
        57039.031 &  \textgreater21.00    &   \_ & \_  & \_  & \_  \\
        57041.026 &  \textgreater20.84    &   \_ & \_  & \_  & \_  \\
         \noalign{\smallskip} 
    \hline
    \end{tabular}
    \caption{Photometry of OGLE14-131 from the OGLE telescope. Magnitudes are observed Vega magnitudes where no extinction correction has been applied.}
    \label{tab:oglephot}
    \end{minipage}
\end{table}

\begin{table} 
\begin{minipage}{\linewidth}
    \begin{tabular}{l l l l}
            \hline
            \noalign{\smallskip}
            MJD & Mag & Error & Filter\\
            \noalign{\smallskip}
            \hline
            \noalign{\smallskip}
    56976.116  &  20.23 &  0.08 &  g$'$\\
    56984.055  &  20.62 &  0.04 &  g$'$\\
    56992.125  &  20.85 &  0.07 &  g$'$\\
    57000.072  &  21.25 &  0.07 &  g$'$\\
    57005.069  &  21.35 &  0.04 &  g$'$\\
    57012.065  &  21.52 &  0.04 &  g$'$\\
    57048.101  &  22.75 &  0.10 &  g$'$\\
    57075.060  &  23.43 &  0.12 &  g$'$\\
    56976.116  &  20.22 &  0.05 &  r$'$\\
    56984.055  &  20.52 &  0.02 &  r$'$\\
    56992.106  &  20.67 &  0.04 &  r$'$\\
    57000.072  &  20.96 &  0.03 &  r$'$\\
    57005.069  &  21.05 &  0.02 &  r$'$\\
    57012.065  &  21.18 &  0.02 &  r$'$\\
    57048.088  &  22.44 &  0.05 &  r$'$\\
    57075.060  &  23.07 &  0.09 &  r$'$\\
    56976.116  &  20.21 &  0.07 &  i$'$\\
    56984.055  &  20.35 &  0.03 &  i$'$\\
    56992.106  &  20.43 &  0.05 &  i$'$\\
    57000.072  &  20.74 &  0.04 &  i$'$\\
    57005.069  &  20.80 &  0.04 &  i$'$\\
    57012.065  &  20.95 &  0.04 &  i$'$\\
    57048.088  &  22.09 &  0.05 &  i$'$\\
    57075.060  &  22.90 &  0.14 &  i$'$\\
    56976.116  &  20.35 &  0.08 &  z$'$\\
    56984.055  &  20.51 &  0.05 &  z$'$\\
    56992.106  &  20.73 &  0.07 &  z$'$\\
    57000.072  &  20.89 &  0.07 &  z$'$\\
    57005.069  &  20.94 &  0.06 &  z$'$\\
    57012.065  &  21.20 &  0.06 &  z$'$\\
    57048.088  &  22.31 &  0.10 &  z$'$\\
    57075.060  &  23.10 &  0.23 &  z$'$\\
    56976.115  &  20.11 &  0.17 &  J\\
    56984.055  &  20.33 &  0.13 &  J\\
    56992.125  &  20.58 &  0.16 &  J\\
    57000.070  &  20.88 &  0.21 &  J\\
    57005.056  &  21.18 &  0.27 &  J\\
    57012.033  &  21.30 &  0.22 &  J\\
    57048.088  &  \textgreater21.72 & \_ & J\\
    57075.060  &  \textgreater21.58 & \_ & J\\
    56976.115  &  \textgreater20.58 & \_ & H\\
    56976.115  &  \textgreater19.88 & \_ & K\\
    \noalign{\smallskip}
    \hline
    \end{tabular}
      \caption{Photometry of OGLE14-131 from GROND. Magnitudes are observed magnitudes, (AB mags for optical, Vega system for NIR) where no extinction correction has been applied. Since there were no detections in the $H$ and $K$ bands, we only include the first non-detection upper-limits from shortly after maximum light. The typical absolute accuracies for GROND photometry are $\pm$0.03~mag in $g^\prime r^\prime i^\prime z^\prime$  and $\pm$0.05~mag in $JHK_{\rm s}$.}
    \label{tab:grondphot}
    \end{minipage}
\end{table}

\clearpage
\begin{table*}
    \centering
    
    \begin{tabular}{llllllllll}
            \hline
            \noalign{\smallskip}
        RA & DEC & $g'$-Mag & $g'$-Error & $r'$-Mag & $r'$-Error & $i'$-Mag & $i'$-Error & $z'$-Mag & $z'$-Error \\
            \noalign{\smallskip}
            \hline
            \noalign{\smallskip}
        018.53470 & -77.10966 & 20.019 & 0.012 & 19.207 & 0.005 & 18.892 & 0.009 & 18.700 & 0.014  \\
        018.46545 & -77.10302 & 20.213 & 0.014 & 19.700 & 0.006 & 19.508 & 0.013 & 19.393 & 0.025  \\
        018.45547 & -77.09666 & 20.762 & 0.019 & 20.188 & 0.010 & 19.942 & 0.018 & 19.743 & 0.035  \\
        018.39682 & -77.10910 & 18.599 & 0.004 & 18.187 & 0.003 & 18.006 & 0.004 & 17.886 & 0.007  \\
        018.48157 & -77.09346 & 18.489 & 0.004 & 17.766 & 0.002 & 17.458 & 0.003 & 17.253 & 0.005  \\
        018.52726 & -77.08878 & 17.923 & 0.003 & 17.534 & 0.002 & 17.378 & 0.003 & 17.293 & 0.005  \\
        018.53799 & -77.09974 & 21.240 & 0.031 & 19.935 & 0.007 & 19.222 & 0.012 & 18.835 & 0.017  \\
        018.57632 & -77.11839 & 19.296 & 0.007 & 19.418 & 0.005 & 19.504 & 0.015 & 19.508 & 0.028  \\
        018.49474 & -77.12227 & 19.827 & 0.010 & 19.330 & 0.005 & 19.120 & 0.010 & 18.985 & 0.019  \\
        018.64939 & -77.11419 & 18.876 & 0.005 & 18.104 & 0.002 & 17.761 & 0.004 & 17.547 & 0.005  \\
        018.64473 & -77.12138 & 20.718 & 0.019 & 19.338 & 0.005 & 18.635 & 0.007 & 18.289 & 0.011  \\
        018.39874 & -77.08181 & 20.916 & 0.025 & 19.545 & 0.006 & 18.439 & 0.005 & 17.911 & 0.009  \\
        018.54510 & -77.07533 & 19.584 & 0.009 & 19.222 & 0.005 & 19.054 & 0.011 & 18.947 & 0.020 \\
        018.39874 & -77.08181 & 19.317 & 0.007 & 18.647 & 0.003 & 18.372 & 0.005 & 18.203 & 0.010 \\
        018.39907 & -77.08963 & 19.951 & 0.011 & 19.391 & 0.005 & 19.156 & 0.011 & 18.986 & 0.017  \\
        018.57524 & -77.09169 & 19.482 & 0.009 & 18.863 & 0.003 & 18.573 & 0.006 & 18.383 & 0.011  \\
        018.65228 & -77.07655 & 19.724 & 0.009 & 19.280 & 0.004 & 19.069 & 0.011 & 18.952 & 0.018  \\
        018.39131 & -77.09921 & 20.893 & 0.024 & 19.607 & 0.006 & 19.012 & 0.009 & 18.714 & 0.015  \\
            \noalign{\smallskip}
            \hline
    \end{tabular} 
    \caption{RA and DEC (J2000); magnitude and error in each of the GROND optical filters, $g'$, $r'$, $i'$, $z'$, of the comparison stars used for calibrating the GROND data.}
    \label{tab:calib}
    
\end{table*}

\end{appendix}

\end{document}